\documentclass[12pt]{article}
\usepackage{amsfonts,amsmath,amssymb,epsf,cite}
\usepackage{graphicx,color}
\usepackage[usenames,dvipsnames]{pstricks}
 \def\m{{\mu}}
 
 \def\n{{\nu}}

 \def\a{{\alpha}}
 
 \def\frac#1#2{{#1\over #2}}

 \def\D{{\Delta}}
 
 \def\s{\sqrt}

 \def\CN{{\cal N}}
 \def\p{\partial}

 \def\al{\alpha'}
 \def\de{\partial}

 \def\f {\frac}
 \def\ti{\tilde}
 \def\ap{\alpha}

 \def\ddd{\cdot\cdot\cdot}
 \def\no{\nonumber \\}
\def\nn{\nonumber \\}
 \def\la{\langle}
 \def\lb{\rangle}

 \def\ov{\overline}
 
 \def\vp{\varphi}

 \def\m{{\mu}}
 
 \def\n{{\nu}}

 \def\a{{\alpha}}
 
 \def\frac#1#2{{#1\over #2}}

 \def\D{{\Delta}}
 
 \def\s{\sqrt}

 \def\CN{{\cal N}}
 \def\p{\partial}

\def\be{\begin{equation}}
\def\ee{\end{equation}}
\def\ba{\begin{eqnarray}}
\def\ea{\end{eqnarray}}
\def\vx{{\vec{x}}}
\def\bu{{\bar{u}}}
\def\bt{{\bar{t}}}

\begin{document}
\begin{titlepage}
\thispagestyle{empty}
\begin{flushright}
UK/10-02\\
UT-10-09\\
IPMU10-0083
\end{flushright}

\bigskip

\begin{center}
\noindent{\Large \textbf
{Probe Branes, Time-dependent Couplings and Thermalization in AdS/CFT}}\\
\vspace{2cm} \noindent{
Sumit R. Das$^{a,b}$\footnote{e-mail:das@pa.uky.edu},
Tatsuma Nishioka$^{c}$\footnote{e-mail:nishioka@hep-th.phys.s.u-tokyo.ac.jp}
and Tadashi Takayanagi$^{a}$\footnote{e-mail:tadashi.takayanagi@ipmu.jp}}

\vspace{1cm}
  {\it
 $^{a}$Institute for the Physics and Mathematics of the Universe (IPMU), \\
 University of Tokyo, Kashiwa, Chiba 277-8582, Japan\\
 $^{b}$Department of Physics and Astronomy, \\
 University of Kentucky, Lexington, KY 40506, USA \footnote{Permanent
   Address}
\\
 $^{c}$Department of Physics, Faculty of Science, \\
 University of Tokyo, Tokyo 113-0033, Japan\\

 }
\end{center}

\vspace{0.3cm}
\begin{abstract}

  We present holographic descriptions of thermalization in conformal
  field theories using probe D-branes in $AdS \times S$
  space-times. We find that the induced metrics on D$p$-brane
  worldvolumes which are rotating in an internal sphere direction have
  horizons with characteristic Hawking temperatures even if there is
  no black hole in the bulk AdS.  The AdS/CFT correspondence applied
  to such systems indeed reveals thermal properties such as Brownian
  motions and AC conductivities in the dual conformal field
  theories. We also use this framework to holographically analyze
  time-dependent systems undergoing a quantum quench,
where parameters in quantum field theories,
  such as a mass or a coupling constant, are suddenly changed.
We confirm that this leads to thermal
  behavior by demonstrating the formation of apparent horizons in the
  induced metric after a certain time.

\end{abstract}
\end{titlepage}
\newpage

\tableofcontents
\newpage

\section{Introduction and Summary}
\hspace{5mm}
The AdS/CFT correspondence \cite{Maldacena,GKP,W,AdSR} has been useful
in understanding time-dependent processes on both sides of the
duality.  On the gravity side, some insight has been obtained for
time-dependent gravitational backgrounds like space-like singularities
using the gauge theory dual \cite{Awad:2009bh,hertog,eva}.  However,
the correspondence has been most successful in the exploration of
non-equilibrium properties of strongly coupled field theories at
finite temperature and/or finite density using their gravity
duals. The significant results in this area include the well known
predictions for linear response functions \cite{SS,viscoref,cmtref1,Ha}.
In recent
years, however, this approach has been extended to the non-linear
domain as well, e.g. in the derivation of non-linear fluid dynamics
from gravity \cite{minwalla} and in the study of thermalization in
terms of the dual process of black hole formation
\cite{janik,Bhattacharyya:2009uu}.

A particularly interesting class of non-equilibrium problems involves
time-dependent couplings or masses which pass through a critical point
either at finite temperature or at zero temperature
\cite{dsenreview}. This problem has experimental relevance in cold
atom physics and has been studied theoretically in two extreme
limits. In the first limit, a parameter is changed quickly through a
critical point (a quantum quench) and the subsequent time evolution is
studied \cite{CCa,CCb,Inte,CCc,CCd,SCa,SCb}. In the other limit, the
parameter changes slowly across a critical point \cite{slowqc}. In
both cases, dynamical quantities have universal signatures of the
critical point. There are, however, few conventional analytical tools
to investigate such problems. It would be clearly interesting if
gauge-gravity duality can be applied to such situations.

In this paper we take a small step in this direction. We investigate
the dual bulk dynamics of a class of
conformal field theories with time-dependent couplings, starting in an
initial vacuum state. In known field theories, this is an interesting
problem even when the couplings do not pass through a critical
point. For example, if we start with a {\em free} massive field theory
in its vacuum state and suddenly change the mass to a different value,
the resulting real time correlation functions become thermal at late
time \cite{CCb,CCc}. Similar behavior can be established in some
interacting field theories as well \cite{CCc,SCb}, but the tools for
investigating such phenomena in strongly coupled theories are rather
limited.

When the field theory has a gravity (as opposed to full fledged
string) dual, a time-dependent coupling corresponds to exciting a {\em
non-normalizable} mode of the dual bulk field and calculating the time
evolution. This is quite similar to the investigation of cosmological
backgrounds in \cite{Awad:2009bh},
forced fluid dynamics \cite{Bhattacharyya:2008ji} or black hole formation
\cite{janik,Bhattacharyya:2009uu}.  However
the class of field theories we consider are considerably simpler, and
may have different applications.  These are defect or flavored
conformal field theories resulting from probe D-branes in a background
space-time of the form of $AdS_m \times S^n$. Such probe branes
\cite{KR} have
been used to model flavor physics \cite{KK,Myers,Erd} and
quantum critical phenomena : this "top-down"
approach \cite{KB,AEKK,JKT,JKST,HPST} is complementary to the
"bottom-up" approach involving the physics of charged black holes in
AdS \cite{cmtref1,Ha,fermisurface,bottomup}. One positive feature of
this top-down approach is that in this
case the dynamics of the probe branes describe the strongly coupled
dynamics of a {\em known} field theory. Furthermore, since the
gravitational background is not altered in the probe limit, the entire
dynamics is given by the worldvolume dynamics of the probe
brane. Solving this dynamics is usually easier than solving Einstein
equations.

In the following, we will find nontrivial time-dependent classical
solutions of several kinds of probe branes which represent
time-dependent couplings of the dual defect conformal field theories. We
find that the induced worldvolume metric on a large class of such
classical solutions have apparent horizons characterized by a
temperature in spite of the absence of black hole horizons in the AdS bulk
metric\footnote{ Here we would like to mention earlier works on the
presence of horizons in the induced metric.  A general analysis
of appearance such horizons on accelerated D-branes has been done in
\cite{Russo}. World-sheet horizons of
F-string has already been discussed in \cite{wa,wb,wc,wd}.  In
\cite{HMT}, the presence of horizons on time-dependent D-branes in
flat space-time has been analyzed. In \cite{ua} horizons on D-branes
from accelerated observers have been studied in pure AdS
space-time.}. In fact, small fluctuations of the worldvolume fields
around these solutions behave like fields in space-times defined by
this induced metric. From the point of view of the dual field theory,
this means that an initial vacuum state can evolve into a thermal
state in the presence of time-dependent couplings.  The emergence of
apparent and event horizons on the worldvolume is somewhat similar to
the emergence of an acoustic horizon for phonons around a supersonic
fluid flow \cite{unruh}. Indeed, this kind of phenomenon can occur in
a large class of non-linear field theories \cite{Barcelo:2001ah}.

\begin{figure}[htbp]
	\centering
	\vspace*{0.5cm}
	\includegraphics[height=5cm]{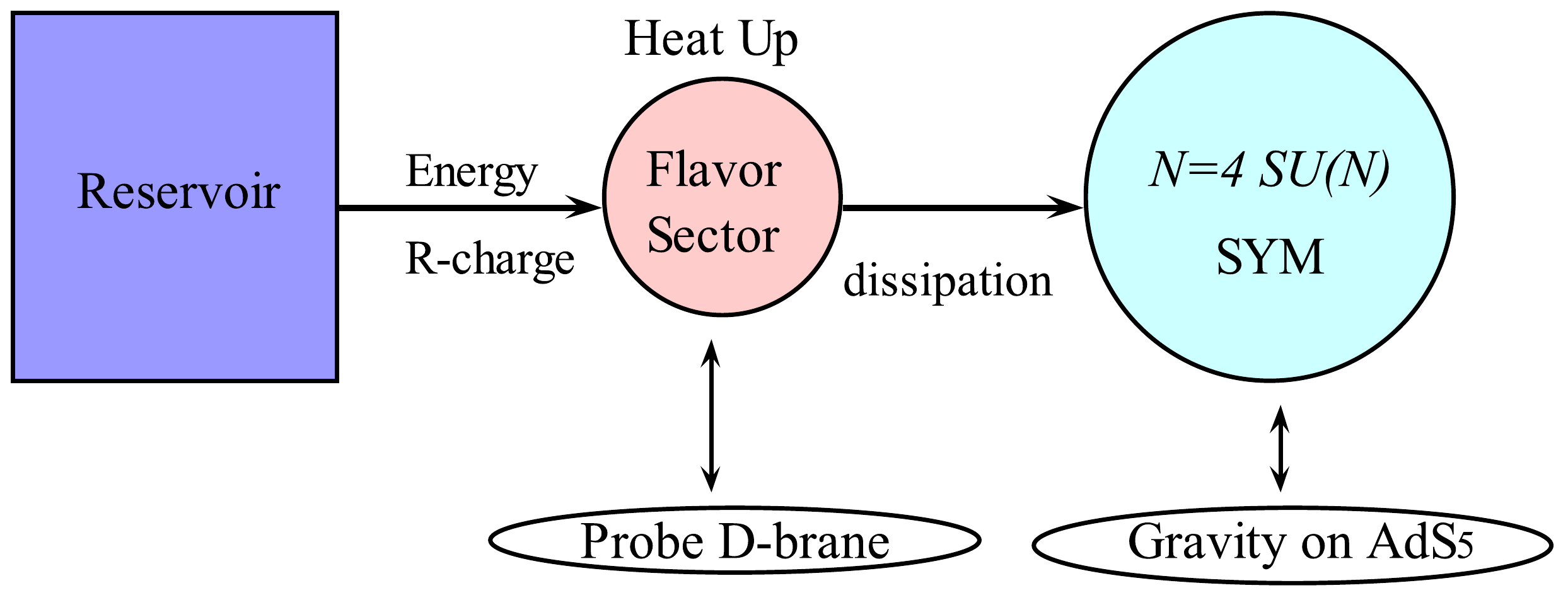}
	\caption{A schematic diagram for our non-equilibrium systems
          in the AdS/CFT.  Here we concentrate on the example of a
          probe D7-brane in $AdS_5\times S^5$ discussed in section
          4. The other cases can be interpreted in the same way after
          obvious modifications.}
	\label{fig:NonEqSystem}
\end{figure}

It is very important that in our probe approximation the dual field
theory should be regarded as an {\it open system}.  Consider such a setup
in $AdS_5\times S^5$ dual to the four-dimensional $\CN=4$ super
Yang-Mills (SYM) which couples to a defect or flavor conformal field
theory (see Fig.\ \ref{fig:NonEqSystem}). Though the energy supplied by the time-dependent coupling
first excites the defect or flavor sector, this energy will finally
dissipate into the $\CN=4$ SYM sector due to the interactions between
each other. In the bulk description this means that energy can flow
from the probe brane to the bulk gravitational degrees of freedom.
Indeed we find that typically the worldvolume apparent horizon evolves
into an event horizon similar to that of a static black hole space-time only when the time
derivative of the classical solution approaches a constant. In such a
case, the long time correlators of fluctuations become thermal with
the Hawking temperature of the event horizon. This can be explicitly
demonstrated for the cases where the fluctuations can be
decoupled. For most cases, these fluctuations are coupled in a
non-minimal way. However, one expects that correlators will still
inherit the thermal behavior implied by the background metric.

In summary, our holographic setups can be regarded as an example of
{\it non-equilibrium steady states}. One of the most important
non-equilibrium properties of our systems is that there are {\it two
different temperatures}: one is the bulk temperature $T_B$ which is
vanishing for pure AdS and the other one is the Hawking temperature
$T_H$ of the apparent horizon of the induced metric of probe D-branes.

The simplest example in our setups is a probe D-string or F-string
extended along the radial direction of $AdS_5$. In the dual $\CN=4$ SYM
theory this represents a monopole or a quark. In this case the
relevant class of classical solutions can be found analytically. In
these solutions, the end-point spins along a direction on the $S^5$
leading to a wave moving along the string towards the Poincare
horizon.  In the dual defect CFT this corresponds to a certain
time-dependent coupling of the hypermultiplet fields with $\CN=4$ vector
multiplet fields \cite{Dia,ooguri}.
This generates a R-charge chemical potential only
for the hypermultiplet and not for the $\CN=4$ SYM sector.  If we start
with a static string in the far past and turn on the spin, an apparent
horizon develops on the worldsheet.  When the value of the spin
approaches a constant at late times, the induced metric coincides with
that of the BTZ metric with an angular coordinate suppressed. The
Hawking temperature of the event horizon is proportional to the final
angular velocity $\omega$, given by simply $T_H=\f{\omega}{2\pi}$. Fluctuations
around this classical solution satisfy simple equations in this
background metric leading to correlators which are thermal.

In particular, this leads to a Brownian motion of the end point.  In
the dual field theory this manifests itself as a Brownian motion of
the monopole position as well as a Brownian motion in the internal
space $SO(6)$. When the string extends all the way to the boundary,
the mass of the monopole or the quark is infinite. However terminating
the string on a suitable space-filling D7-brane makes the mass finite. It
should be again emphasized that this is a {\em non-equilibrium}
situation : it is only the monopole or quark which behaves thermally
while the other $\CN=4$ degrees of freedom are still at zero
temperature. $1/N$ corrections will eventually thermalize the latter
- but that requires going beyond the probe approximation.  Since we
have an {\em open} system, energy can flow into the vector multiplet
sector. In a closed system, we typically expect thermal behavior if
the time dependence {\em vanishes} at late times. In contrast, in our
case we need a steady time dependence to achieve thermality since
energy can flow out of the defect CFT at a constant rate.  Such
Brownian motion behavior has been observed for fluctuations around
static strings in the background of AdS black holes, i.e. when the
field theory is already at some finite temperature \cite{Bra,Brb}. In
our case, the parent field theory is at zero temperature - an
effective temperature is produced by a time-dependent coupling.

The physics of other probe D-branes is quite similar, but requires a
combination of analytical and numerical tools. For D3, D5 and
D7-branes in the $AdS_5 \times S^5$ geometry, we find classical
solutions with angular momenta dual to R-charges which lead to induced
worldvolume metrics with horizons. For a probe D3-brane, which is dual
to a ($2+1$)-dimensional defect CFT, we calculate the conductivity by
turning on a worldvolume gauge field. There is a finite AC
conductivity which arises from dissipation into the $\CN=4$ degrees of
freedom. A similar behavior has been observed earlier using probe
D-branes at finite temperature, i.e. in AdS
black hole backgrounds \cite{KB,HPST}. In contrast, we observe this
behavior in a zero temperature field theory where a time-dependent
coupling effectively produces a thermal state. In the D7-brane case,
we can rotate a D7-brane in the standard D3-D7 system \cite{KK} dual
to four-dimensional $\CN=4$ SYM with a massive flavor
hypermultiplet. This setup has been already discussed in \cite{Bannon}
(see also \cite{Evans} for a similar analysis for D3-D5 system).
This rotation excites a R-charge only in the
hypermultiplet sector. We find that for a sufficiently large R-charge,
the induced metric of the probe D7-brane has a horizon which surrounds
a cusp-like singularity on the worldvolume. This structure looks very
similar to the structure of the standard black holes in general
relativity. As a final example, we holographically describe a
time-dependent mass in the defect CFT sector dual to a probe
D5-brane. We numerically show that an apparent horizon is formed under
a time-dependent change of mass, thus describing thermalization of the
dual CFT. In free field theory, it is known that the effective
temperature after the quench depends on the momenta of each modes
since they are decoupled with each other in free field theory
\cite{CCc,CCd}.  However, our holographic result predicts that this
picture will be completely changed for the strongly interacting gauge
theories. In the latter case {\it the different momentum modes observe
  a common temperature}.

It is also an intriguing problem to associate an entropy to such a
D-brane with thermal horizon. We will show that one holographic
candidate for such a entropy is divergent in general except in the
lowest-dimensional case i.e. D1-brane case. We suggest that this might
be interpreted as an entanglement entropy when we trace out the $\CN=4$
SYM sector, though we would like to leave its detail for future work.

This paper is organized as follows: In section 2, we consider probe
D$p$-branes which are extended in AdS space-time and rotating in a sphere
direction. We show that an apparent horizon is always formed in the
Poincare AdS space.  We calculate the energy dissipation in such
systems. In section 3, we study the thermal property of the rotating
D$p$-brane solutions constructed in section 2. In particular, we will
study Brownian motion and thermal correlation functions for a
probe D1-brane and the AC conductivity for a D3-brane. In section 4,
we analyze the rotating D7-brane with a mass for flavor
hypermultiplet. We will show that for a sufficiently fast rotation,
the induced metric of the probe D7-brane has a horizon.  In section 5,
we will give a brief analysis of a holographic dual of a quantum quench
caused by a rapid change of a mass parameter in a defect CFT. We consider a
particular case of probe D5-brane and numerically show the appearance
of an apparent horizon in the time-dependent induced worldvolume metric.

\section{Worldvolume Horizons and
Mini Black Holes from Rotating D-branes in AdS}
\hspace{5mm}
Typical setups of AdS/CFT correspondence are $AdS_{d+2}\times X^q$
backgrounds in string theory or M-theory, where $AdS_{d+2}$ is the
($d+2$)-dimensional AdS space in Poincare coordinates and
$X^q$ is a $q$-dimensional
Einstein manifold \cite{Maldacena,GKP,W,AdSR}. In this paper we mainly assume the simplest case
$X^q=S^q$. We can write their metric as follows (we set the AdS radius $R=1$ throughout this paper, for simplicity of
presentation.)
\be
ds^2=-r^2dt^2+r^2\sum_{i=1}^d dx_i^2+\f{dr^2}{r^2}+(d\theta^2+\sin^2\theta d\vp^2+\cos^2\theta
d\Omega^2_{q-2}) \ .\label{adsfi}
\ee
This clearly includes the celebrated example of the $AdS_5\times S^5$ in type IIB string theory
by setting $d=3$ and $q=5$, which is dual to the four-dimensional $\CN=4$ super Yang-Mills theory.

The D-brane systems we would like to consider now are D$p$-branes
rotating in the $S^q$ directions. These rotations are dual to non-vanishing
 R-charges in CFT$_{d+1}$. The equations of motion require that such
D-branes should rotate along the largest circles in  $S^q$. Below we will show that
such D-brane systems can be regarded as mini black holes in AdS spaces and they describe
locally thermal vacua in a probe limit.

Since the analysis of rotating D1-branes is much simpler we will first study this and then proceed to
the analysis of general D$p$-branes.

\begin{figure}[h]
\centering
\vspace{0.5cm}
\includegraphics[height=10cm]{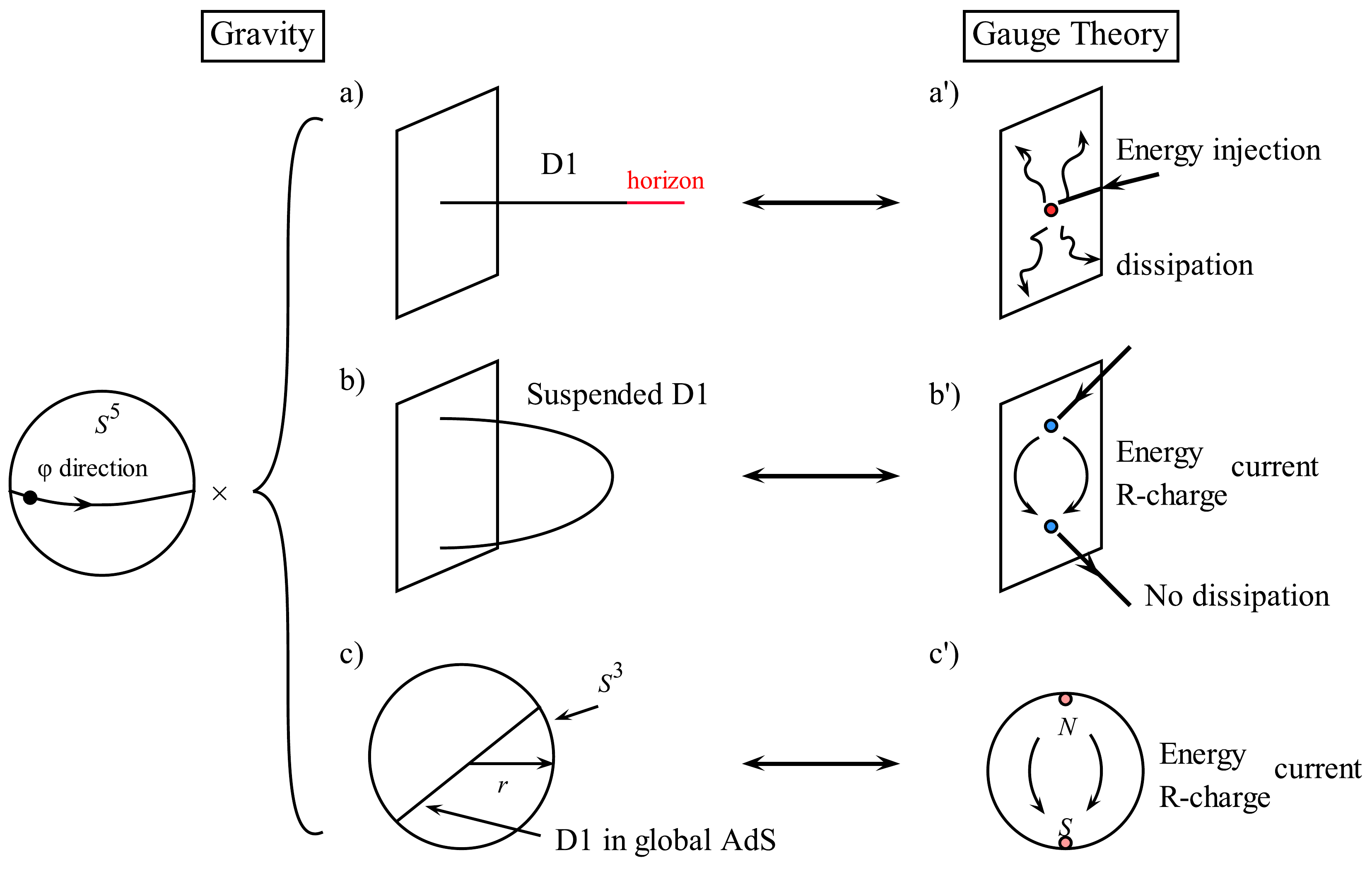}
\caption{We present the schematic pictures of (a) a thermal D1-brane
  solution extended in Poincare $AdS_5$, (b) a suspended D1-brane
  solution in Poincare $AdS_5$, (c) a D1-brane solution in global
  $AdS_5$, and their CFT duals: (a$^\prime$),(b$^\prime$) and
  (c$^\prime$).  The D1-branes are all rotating along a great circle
  in the $S^5$.}
\label{fig:D1branes}
\end{figure}

All the rotating solutions that will be constructed in this section and their relation to dual
gauge theory are summarized in Fig.\ \ref{fig:D1branes}.

\subsection{Rotating D1-branes}
Consider a probe D1-brane (or a F-string equivalently owing to S-duality)
in $AdS_{d+2}\times S^q$ (\ref{adsfi}).
The D1-brane extends in both $t$ and $r$ direction, infinitely and localized at $x_i=0$. In addition, it is spinning
in the $\vp$ direction. Therefore, the D1-brane world-volume is specified by the function
\be
\vp=\vp(t,r),
\ee
at constant $\theta(\equiv\theta_*)$.

In this setup, the DBI action of
the D1-brane is
\ba
&& S_{DBI}=-T_{D1}\int dtdr L \ , \\
&& \ \ L\equiv \s{1+r^2(\vp')^2\sin^2\theta_*-\f{\sin^2\theta_*}{r^2}\dot{\vp}^2} \ ,\label{dbif}
\ea
where $T_{D1}=\f{1}{2\pi\al g_s}$ is the tension of a D-string.
The equation of motion reads
\be
\f{\de}{\de r}\left(\f{r^2\sin^2\theta_* \vp'}{L}\right)=\f{\de}{\de t}\left(\f{\sin^2\theta_* \dot{\vp}}{r^2 L}\right).
\ee
Consider solutions of the form
\ba
&& \vp=\omega t +g(r) \ ,\ \ \ \ \  g'(r)=\s{\f{1-\omega^2/r^2}{A r^4-r^2}} \ ,\\
&& \ \  \theta_*=\f{\pi}{2} \ ,
\ea
where $A$ is an arbitrary constant. The Lagrangian density for this solution is
\be
L=\s{\f{A(r^2-\omega^2)}{Ar^2-1}} \ .
\ee
In order to avoid a negative sign inside the square root, we need to require
\be
A=\f{1}{\omega^2} \ .\label{relationa}
\ee
In this way, we find the rotating D1-brane is given by the solution
\be
\vp=\omega t-\f{\omega}{r}+\vp_0 \ , \label{done}
\ee
up to an integration constant $\vp_0$.

The induced metric on this D1-brane worldsheet is
\ba
ds^2&=&-(r^2-\omega^2)dt^2+2\f{\omega^2}{r^2}dtdr+\left(\f{1}{r^2}+\f{\omega^2}{r^4}\right)dr^2.
\ea
We can rewrite this by defining a new time coordinate
\be
\tau\equiv t-\f{1}{r}-\f{1}{2\omega}\log\f{r-\omega}{r+\omega} \ ,
\ee
as follows
\be
ds^2=-(r^2-\omega^2)d\tau^2+\f{dr^2}{r^2-\omega^2} \ .  \label{d1met}
\ee
Notice that $\tau$ becomes identical to $t$ as we approach the boundary $r\to \infty$.

Interestingly, the induced metric (\ref{d1met}) coincides with the BTZ black hole
with the angular coordinate suppressed. Therefore it has a horizon at $r=\omega$ and
we can read off its Hawking temperature $T_H$
\be
T_{H}=\f{\omega}{2\pi} \ ,  \label{d1temp}
\ee
by Wick-rotating $\tau$ into a Euclidean time (see the picture (a) in Fig.\ \ref{fig:D1branes}).

If we take into account the back-reaction of this solution to the supergravity background, it is natural to
expect that such D1-branes produce a very small black hole in the bulk AdS localized in the
$R^{d}$ direction (we may call it a mini black hole).

This shows that the rotating D1-brane describes a thermal object with
temperature $T_H$ in the dual CFT .  Notice that the bulk of AdS is at
zero temperature. If we concentrate on the most important example of
$AdS_5\times S^5$, our system is dual to $\CN=4$ super Yang-Mills
theory coupled to an infinitely heavy monopole. The $\CN=4$ gauge
theory itself is at zero temperature, while the monopole is at finite
temperature $T_H$. Therefore such systems are in non-equilibrium
steady states.

The rotating D$1$ brane solution corresponds to a time dependent
coupling in the $\CN =4$ theory coupled to hypermultiplets living on the
zero dimensional defect.  The D1-D3 system is 1/4-BPS and the D1-D3
open strings lead to the two complex scalars $(Q,\ti{Q})$ of
hypermultiplets which belong to the fundamental and anti-fundamental
representations of the color $SU(N)$ gauge group. Let us
express the three complex adjoint scalar fields
in the $\CN=4$ super Yang-Mills by $(\Phi_1,\Phi_2,\Phi_3)$. These
correspond to cartesian coordinates in the transverse $C^3$ composed of
$(r,\Omega_5)$ where $\Omega_5$ represents the 5-sphere.  We choose
$\Phi_3$ such that its phase rotation describes the one in the $\vp$
direction and that $\theta=\pi/2$ is equivalent to
$\Phi_1=\Phi_2=0$. We now argue that the time dependent coupling term
which corresponds to a uniformly rotation D1-brane is given by
\be \int dt
\left[\ov{Q}~ \left[\mbox{Im}(\Phi_3 e^{-i\omega t})\right]^2~Q
  +\ti{Q}~\left[ \mbox{Im}(\Phi_3 e^{-i\omega
      t})\right]^2~\ov{\ti{Q}}\right] .
\label{intti}
\ee
To see this consider going to the Coulomb branch where $\langle \Phi_3\rangle \neq
0$. This corresponds to separating out some number $k$ of the original
$N$ D3-branes. Apart from generating masses for the off diagonal
vector multiplet fields, the coupling in (\ref{intti}) generates
masses for appropriate components of the hypermultiplet
fields. Consider the simplest situation where only one D3-brane is
separated in the $\theta = \pi/2$ hyperplane, so that we only have
$\langle \Phi_3\rangle = {\rm diag} [\Phi_3,0,0,\cdots]$.
In this case the topmost component of $Q$ would become
massive. This mass should be equal to the length of the shortest open
string joining the D1-brane with this separated brane (upto a factor
of the string tension). The rotating D1 described above is described
by $(0,0,ze^{i\vp(t)}=ze^{i\omega t})$ in the transverse $C^3$,
where $z$ takes any real
value.  The shortest distance is then given by
$|\mbox{Im}(\Phi_3 e^{-i\omega t})|$, leading to a mass of
the
scalar fields $(Q,\ti{Q})$
exactly as predicted by
(\ref{intti}). In the above discussion we have taken
$\langle \Phi_1\rangle = \langle \Phi_2\rangle =0$ to extract
the main time-dependent part of the interaction. However the result
can be generalized to a generic point in the Coulomb branch, where
other terms involving the $\Phi_1$ and $\Phi_2$ would be involved.

As a further check on consistency, note that the operator in
(\ref{intti}) has dimension $1$. As we will see in a later section the
field $\vp$ is a massless minimally coupled scalar on the
worldsheet. The usual AdS/CFT formula (now applied to $AdS_2$ on which
the D1-brane wraps) relating bulk masses to conformal
dimensions then predict that the dual operator should have dimension
$1$.

We thus see that
the AdS/CFT correspondence shows that if we consider a system with such
time-dependent interactions, the system gets thermal (see Fig.\
\ref{fig:NonEqSystem}).

\subsection{General F1/D1 solutions}

It turns out that there are more general purely right-moving or purely left-moving solutions of D1
or F1 branes in the background of the class of metrics which include $AdS \times S$ as well as
 black branes or black holes in Poincare or global coordinates. Consider a metric of the form
\be
ds^2=-f(r)dt^2+\f{dr^2}{f(r)}+r^2\sum_{i=1}^d dx_i^2+(d\theta^2+\sin^2\theta d\vp^2+\cos^2\theta
d\Omega^2_{q-2}) \ , \label{metgl}
\ee
and introduce the outgoing and ingoing Eddington-Finkelstein coordinates
\be
u = t-\int \frac{dr}{f(r)} \ ,~~~~~~~v = t + \int \frac{dr}{f(r)}~.
\label{uvdef}
\ee
The DBI action then becomes
\be
S = -T_{D1} \int du dv~f(r) \left[1-\frac{4}{f(r)}\partial_u \vp \partial_v \vp \right]^{1/2} \ .
\label{uvdbi}
\ee
The equation of motion then becomes
\be
\partial_u\partial_v \vp + \frac{2}{L} \partial_v \vp \partial_u \left( \frac{\partial_u \vp \partial_v \vp}{f(r)} \right)
+ \frac{2}{L} \partial_u \vp \partial_v \left( \frac{\partial_u \vp \partial_v \vp}{f(r)} \right) = 0 \ ,
\label{uveqnmotion}
\ee
where
\be
L = 1-\frac{4}{f(r)}\partial_u \vp \partial_v \vp \ .
\ee
It is easy to see that any $\vp$ which satisfies either $\partial_u \vp = 0$ or $\partial_v \vp = 0$ is a solution of (\ref{uveqnmotion}). In
particular, solutions which are functions of $v$ alone may be thought of the retarded effect of a boundary value of $\vp$.

The induced metric produced by such a retarded solution is given by
\be
\label{indf} ds_{ind}^2 = -f(r) dudv +(\partial_v \vp)^2 dv^2 =
2drdv - [ f(r)-(\partial_v \vp)^2]dv^2 \ .
\ee
This is a two-dimensional AdS Vaidya metric which shows the formation of an apparent
horizon at $f(r) = (\partial_v \vp)^2$, provided this equation has a solution
for real $r$.  The dual CFT description is
captured by the interaction (\ref{intti}) with $\omega t$
is replaced with the general function $\vp(t)$ as is clear from the derivation in (\ref{intti}).

Special cases of the background metric (\ref{metgl}) include Poincare patch $AdS_5$, with $f(r) = r^2$ and black 3-branes with $f(r) = r^2(1-\frac{r_0^4}{r^4})$. With the replacement of $\sum_i(dx_i)^2 \rightarrow d\Omega_3^2$ this can include global $AdS_5$ as well as AdS-Schwarzschild. We will discuss these solutions further in the following sections.

In the special case where the function $\vp (v)$ approaches
$\vp (v) \rightarrow \omega v$ at late retarded times, the apparent horizon asymptotes to an event horizon at $f(r) = \omega^2$ which signals thermalization.
In dual CFT, this system consists of two sectors with different temperatures.

\subsection{Rotating D$p$-branes}\label{Dpp}

We can extend the previous system to the rotating higher-dimensional D$p$-branes.
Consider a rotating D$p$-brane for $p\le d+1$ in $AdS_{d+2} \times S^q$ space-time which extends in $(t,r, x_{1}\dots, x_{p-1})$
directions of AdS and rotates in $\varphi$ direction at $\theta =\pi/2$.
We again assume $\varphi$ depends on only $t$ and $r$ directions.

The DBI action is given by
\begin{align}
	S_{DBI} = - T_{p}\int dtdr d^{p-1}x ~ r^{p-1} \s{1 + 
	(r^{2}\varphi'^{2} - \f{\dot\varphi^{2}}{r^{2}}) } \ .
\end{align}
It follows that we obtain the equation of motion:
\begin{align}
	\p_{t} \left( \f{r^{p-3}\dot\vp}{\s{1+(r^{2}\varphi'^{2} - \f{\dot\varphi^{2}}{r^{2}})}} \right) = \p_{r}
	\left( \f{r^{p+1}\vp'}{\s{1+(r^{2}\varphi'^{2} - \f{\dot\varphi^{2}}{r^{2}})}} \right) \ .
\end{align}
We now consider the rotating solution with angular velocity $\omega$
along $\vp$ direction, and we thus assume the following
form of the solution:
\begin{align}
	\vp (t,r) = \omega t + g(r) \ . \label{dpsolone}
\end{align}
Substituting the ansatz into the equation of motion the function $g(r)$ looks like
\begin{align}
	g' = \s{\f{A^{2}(r^{2}-\omega^{2})}{r^{4}(r^{2p}-A^{2})}} \ , \label{dpsoltwo}
\end{align}
where $A$ is the integration constant, and should be set to $A=\omega^{p}$ to
avoid a divergence of the function $g(r)$ at $r=\omega$.

The induced metric on the rotating D$p$-brane is
\begin{align}
	ds^{2}_{ind} &= - (r^{2}-\omega^{2})dt^{2} + (\f{1}{r^{2}} + g'^{2})dr^{2} + 2\omega g' dtdr + r^{2}d\vec x_{p-1}^{2} \no
	&= - (r^{2}-\omega^{2}) d\tau^{2} + \f{r^{2(p-1)}}{r^{2p}-\omega^{2p}}dr^{2} + r^{2}d\vec x_{p-1}^{2}  \ ,\label{genind}
\end{align}
where the new time coordinate $\tau$ is defined as
\begin{align}
	\tau = t - \int dr\f{\omega^{p+1}}{r^{2}\s{(r^{2p}-\omega^{2p})(r^{2}-\omega^{2})}} \ .
\end{align}
This metric (\ref{genind}) clearly has a horizon at $r=\omega$.
After Wick rotation of $\tau$ we can determine its Hawking temperature by
demanding smoothness at $r=\omega$:
\begin{align}
	T = \f{\s{p}\,\omega}{2\pi} \ .
\end{align}

\subsection{Back-reactions and Charge/Energy Dissipations}

In the above analysis, back-reaction to the supergravity sector has been neglected since we employed the
probe approximation. We would now like to see how much and when this is justified. Since the essential
issues are the same, we concentrate on a rotating D1-brane located at $x_i=0$.

\begin{figure}[htbp]
 \begin{center}
      \includegraphics[width=7cm]{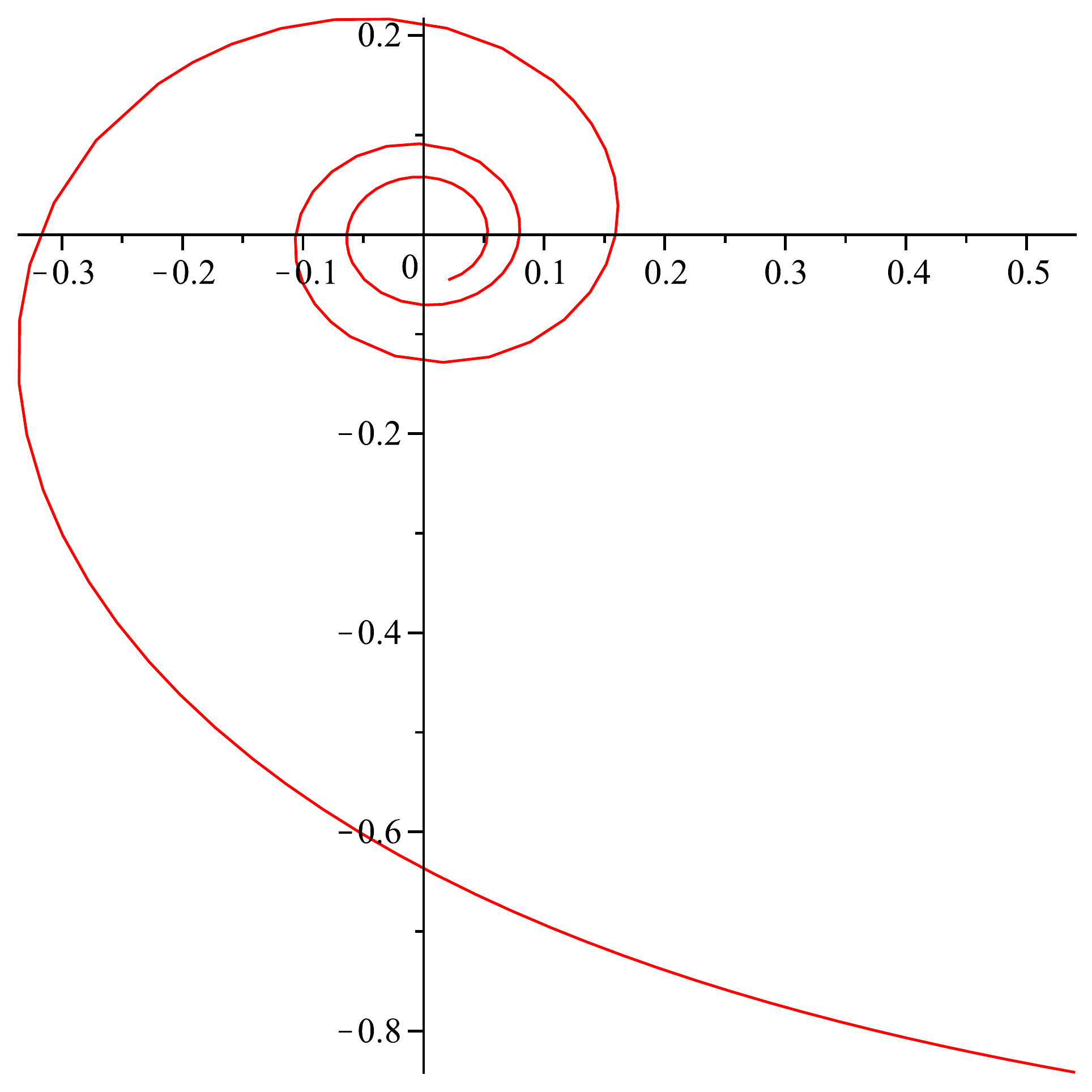}
		\end{center}
\caption{The profile of a rotation D1-brane in AdS at some given time in the
$(x^4=r \sin \vp, x^5=r\cos \vp)$ plane. The time evolution
is obtained by rotating the profile.}
\label{profile-d1}
\end{figure}

The total energy of our rotating D1-brane conjugate to time $t$ is
given by
\be
E=\int^\infty_0
dr\f{1+r^2\vp'^2}{\s{1+r^2(\vp')^2-\f{1}{r^2}\dot{\vp}^2}}=\int^\infty_0
dr \left(1+\f{\omega^2}{r^2}\right) \ . \label{Ham}
\ee
This shows
that the energy density blows up at $r=0$. Thus in the IR limit we
cannot neglect the gravitational back-reaction. This is intuitively
understood if we plot the form of D1-brane profile (\ref{done}) at
some given time in the $(r\sin\vp,r\cos\vp)$ plane, as in
Fig.\ \ref{profile-d1}. Notice that the D1-brane wraps
infinitely many times at $r=0$. Similar behavior is present for the
rotating D-brane in AdS black holes.

It is reasonable to conclude that the large back-reaction will result in the formation
of a black hole in the bulk of AdS,
centered at $x_i=r=0$. The size of this black hole should grow as the energy is pumped into
it from the D1-brane constantly. To obtain this energy flux we need to calculate the energy-stress tensor \cite{KOT} for the probe D1-brane.
We define the energy-stress tensor by
\be
T^a_{\ b}=\f{2}{\s{-g}}\f{\delta S}{\delta g_{ac}}g_{cb} \ ,
\ee
where $g_{ab}$ denotes the bulk metric (the index $a$ and $b$ runs
all ten (or eleven) coordinates of our space-time).
This satisfies the equation of motion
\be
\nabla_{a}T^a_{\ b}=0 \ , \label{est}
\ee
which is reduced to
\be
\de_{a}(T^a_{\ t} \s{-g})=0 \ ,
\ee
for static space-times. This leads to the conserved energy
\be
E=\int dr \s{-g}\ T^t_{\ t} \ .
\ee

In our rotating D1, the explicit expressions for $T^a_t$ are
\ba &&
\s{-g}\ T^t_{\ t}=T_{D1}\f{1+r^2\vp'^2}{\s{1+r^2\vp'^2-\dot{\vp}^2/r^2}}=T_{D1}
\left(1+\f{\omega^2}{r^2}\right) \ ,\no &&
\s{-g}\ T^r_{\ t}=T_{D1}\f{r^2\dot{\vp}\vp'}{\s{1+r^2\vp'^2-\dot{\vp}^2/r^2}}=T_{D1}\omega^2 \ ,\no
&&
\s{-g}\ T^{r}_{\ r}=-T_{D1}\f{1-\dot{\vp}^2/r^2}{\s{1+r^2\vp'^2-\dot{\vp}^2/r^2}}=-T_{D1}
\left(1-\f{\omega^2}{r^2}\right) \ .   \label{esth}
 \ea
Thus we can calculate the time evolution of the total energy as
\be
 \f{dE}{dt}=\f{d}{dt}\int dr \s{-g}T^t_{\ t}=\int dr \de_r(\s{-g}T^r_{t})=\s{-g}~T^r_{t}~\big|^{r=\infty}_{r=0}
=T_{D1}\omega^2-T_{D1}\omega^2=0 \ . \label{endi} \ee
Even though the total energy does not depend on time, this result shows that
the energy $\f{dE}{dt}=T_{D1}\omega^2$ per unit time is injected at the boundary $r=\infty$ by some external system
(reservoir) and the same energy $\f{dE}{dt}=T_{D1}\omega^2$ is dissipated from the IR region into the bulk AdS
as summarized in the picture (a$^\prime$) in Fig.\ \ref{fig:D1branes}.
This dissipation from the D1-brane to the bulk AdS will create a localized black hole in AdS. The dissipation via the AdS horizon is very similar to the phenomenon
called drag force, which is dual to the jet quenching \cite{Gu,HKKKY,LRW}.

One may wonder how we can explicitly understand the external injection of energy in our stationary solutions
as we argued. To see this, let us introduce UV and IR cut off into our rotating D1-brane system so that
it extends from $r=r_{IR}$ to $r=r_{UV}(\gg r_{IR})$. In this setup
 it is clear that the energy flux $T_{D1}\omega^2$
is coming from $r=r_{UV}$ and the same energy is going out at $r=r_{IR}$. The presence of this energy flux is
clear from the non-vanishing value of $T^r_t$ in (\ref{esth}) at $r=r_{IR}$ and $r=r_{UV}$. At $r=r_{IR}$, the energy is not reflected back but its back-reaction will make a localized black hole, which absorbs the injected
energy. Another way to see the energy flow explicitly
is to consider general time-dependent solutions presented in section 2.2. In this case we can choose the
D1-brane profile such that the energy is injected like a pulse and this travels into the IR region.
In this case, $T^r_t$ gets non-vanishing at the UV cut off only during a certain time interval.
Our stationary rotating D1-brane solution can be regarded as an infinitely many collections of such pluses
and therefore it requires the constant energy injection at the AdS boundary.

Similarly we can calculate the dissipation of angular momentum which is dual to a R-charge in the CFT.
The angular momentum of a rotating D1 is given by
\be Q_{R}=\f{\delta S}{\delta \dot{\vp}}=T_{D1}\int dr
\f{\omega}{r^2} \ ,\ee which is divergent like its energy.

The emission toward the horizon $r=0$ for the Poincare AdS can be
estimated as
\be \f{d Q_R}{d t}\Biggr|_{r=0}=-T_{D1}\f{r^2\vp'}{\s{1+r^2\vp'^2-\dot{\vp}^2/r^2}}\Bigr|_{r=0}
=-T_{D1}\omega \ . \ee This amount of angular momentum (=R-charge)
comes from the boundary $r=\infty$ and flows into the horizon $r=0$.
For higher-dimensional rotating D$p$-branes, the situations are almost the same.
{}From the viewpoint of the boundary CFT which lives on $R^{1,d}$, only the codimension $d-p+1$ part, defined by the intersection between the AdS boundary and the D$p$-brane, is thermalized at the temperature $T_H$, while the other part is at zero temperature in the probe approximation.
The energy dissipation we found in (\ref{endi}) corresponds to the energy flow and the R-charge flow from the thermal part to the cold region in the bulk of $R^{1,d}$. Therefore the temperature of the bulk region should constantly increase. This is dual to a growing black hole once back-reaction is taken into account.


\subsection{Rotating D-branes in Global AdS}

One way to suppress a large IR back-reaction is to replace the Poincare $AdS_{d+2}$ with
the global $AdS_{d+2}$. Again we will concentrate on D1-branes just for simplicity.
Then we need to take $f(r)=r^2+1$ in the equation (\ref{metgl}).
The rotating D1-brane is described by the profile \be \vp=\omega (t
+\arctan r) \ . \label{prgl} \ee
Notice that the D1-brane in the global AdS pierces the center of the AdS and thus
has two end points at the boundary, which are
identified with the north-pole and south-pole of the boundary $S^3$, respectively, as shown in
the picture (c) in Fig.\ \ref{fig:D1branes}. This is conveniently
described by extending the range of $r$ in (\ref{prgl}) to $-\infty<r<\infty$.

The induced metric on the worldsheet is given by
\be
ds^2_{ind}=-(r^2+1-\omega^2)d\tau^2+\f{dr^2}{r^2+1-\omega^2} \ ,
\ee
where
$\tau=t+h(r)$ and $h'(r)=\f{\omega^2}{(1+r^2)(r^2+1-\omega^2)}$.
Therefore, a horizon exists only when $|\omega|>1$. This gap is analogous to the
Hawking-Page transition between the global AdS and the AdS black hole \cite{WiT}.
In our case,
it is dual to the confinement/deconfinement phase transition
of the hypermultiplets from D3-D1 strings. The Hawking temperature is given by
$T_H=\f{\s{\omega^2-1}}{2\pi}$ \ .

The total energy of the D1-brane is given by
\be E=T_{D1}\int^\infty_{-\infty} dr
\left(1+\f{\omega^2}{r^2+1}\right). \ee Its divergent part which is independent of
$\omega$ can be canceled by the standard holographic renormalization.
The total angular momentum becomes
finite for the D1 in global AdS as given by \be Q_{R}=T_{D1}\int^\infty_{-\infty}  dr
\f{\omega}{r^2+1}=\pi T_{D1}\omega \ .\ee
Thus we can conclude that the back-reaction of the probe D1-brane to the
metric is not large even when $\omega$ is non-zero.
This is rather different from the rotating D1-brane in Poincare AdS.

Just as in the Poincare patch solutions in the previous sections, there are general purely left moving or purely right moving solutions. The solutions corresponding to the retarded effect of specified boundary conditions are of the form
\be \vp=\vp(t+\arctan r) \ , \ee
for arbitrary choice of the function $\vp(v)$.
The R-charge current shows the behavior
\be
\f{dQ_R}{dt}=(r^2+1)\vp'|^{\infty}_{-\infty}=\dot\vp(t+\pi/2)-\dot\vp (t-\pi/2)\equiv j_1-j_2 \ .
\ee
This means that in the dual CFT, the current $j_1$ starts from the north pole $P_1$, it flows into the
bulk of $S^3$ and is eventually absorbed at the south pole $P_2$ as the current $j_2$ (see the picture (c$^\prime$) in Fig.\ \ref{fig:D1branes}). The time delay $\Delta t=\pi$ is just the
propagation of this current at the speed of light from $P_1$ to $P_2$. Notice that in this global
AdS, there is no energy dissipation toward the bulk (i.e. $\CN=4$ SYM sector) at planar order, which
will be because the bulk system is in the confinement phase.
The current is carried by flowing R-charged particles. Also notice that the effective
temperature for $P$ and $Q$ are $\f{j_1}{2\pi}$ and $\f{j_2}{2\pi}$.

\subsection{Suspended Rotating D1-branes in Poincare AdS}

As we have seen, in the global AdS space, the rotating D1-brane always has two end points
at the boundary. This is because we cannot have a single source of charge current in a compact
space. In the Poincare AdS case, this is no longer true since the space is non-compact. However,
still it is interesting to consider a system where there are two point sources $P_1$ and $P_2$
in the boundary $R^{1,3}$. In this case, there are
two possibilities of rotating D1-branes. The first one is two disconnected rotating D1-branes
infinitely stretching in the radial direction. As we already analyzed, this has a thermal horizon
at $r=\omega$ and has the temperature $T=\f{\omega}{2\pi}$. We can take the temperatures of two points
independently, denoted by $T_1$ and $T_2$. Then the total R-charge current flow from
$P_1$ to $P_2$ in the dual CFT is given by
\be
\f{dQ_R}{dt}=2\pi T_{D1}(T_1-T_2) \ , \label{cur}
\ee
and the energy flow is
\be
\f{dE}{dt}=4\pi^2T_{D1}(T_1^2-T_2^2) \ .\label{ecur}
\ee
At the two points, the system gets thermalized and
the carrier possesses the energy $\omega$ per a unit R-charge.
They will strongly interact with the bulk gauge theory even though
it is at zero temperature. Notice that in this case there is no correlation between two
points and the physics does not depend on the distance $L$ between them.

We can calculate the R-charge conductance from (\ref{cur}). We consider the
setup of type IIB string on $AdS_5\times S^5$ and recover the radius of the AdS space
$R=\left(4\pi g_sN \al^2\right)^\f{1}{4}$. The chemical potential
at the two points are identified with $\omega_1$ and $\omega_2$. Thus we find the conductance,
which is the ratio of voltage $V$ to the current $I$, is given by
\be
\f{I}{V}=\f{\dot{Q_R}}{\omega_1-\omega_2}=T_{D1}R^2=\f{1}{\s{\pi}}\f{N}{\s{\lambda}} \ ,
\ee
where $\lambda=N g_s$ is the 't Hooft coupling of the dual $\CN=4$ SYM. Notice that we here
normalized the R-charge such that a transverse complex scalar $\Phi_3$ in the $\CN=4$ has $+1$
i.e. $R(\Phi_3)=1$. For (1+1)-dimensional fermi liquids, the value of the conductance
is quantized into $\f{{\cal T}}{2\pi}$ times integers (${\cal T}$ is the transmission probability),
known as the Landauer formula \cite{La}.
On the other hand, in our case, it is quantized in the unit of $\f{1}{\s{\lambda}}$
and we may estimate ${\cal T}\sim \f{1}{\s{\lambda}}$ for the $\CN=4$ SYM media.

The second solution is a suspended rotating D1-brane which connects
the boundary two points $P$ and $Q$ as depicted as the picture (b) in Fig.\ \ref{fig:D1branes}.
We describe its profile
by
\ba
&& \vp=\omega t+g(r) \ ,\no
&& x\equiv x_1=x(r) \ .\label{xx}
\ea
 The action looks like
 \be
 S_{DBI}=-T_{D1}\int dt dr\s{K} \ ,
 \ee
 where
 \be
 K=\f{r^2-\omega^2}{r^2}+r^2(r^2-\omega^2)x'^2+r^2g'^2 \ .
 \ee
Using the symmetry under constant translations of $x$ and $g$,
we can simply reduce the
equation of motion into
\ba
g'(r)=\f{B(r^2-\omega^2)}{r^2\s{B^2(r^2-\omega^2)(A^2r^2-1)-1}} \ ,\no
x'(r)=\f{1}{r^2\s{B^2(r^2-\omega^2)(A^2r^2-1)-1}} \ ,
\ea
where $A$ and $B$ are integration constants.

Define $r_*$ as a solution to $B^2(r^2-\omega^2)(A^2r^2-1)-1=0$.
The divergence of $x'$ at $r=r_*$ shows that $r=r_*$ is the turning point of the suspended
D1-brane. Thus the D1-brane starts from $r=\infty$ and turns around $r=r_*$ and gets back to
the boundary $r=\infty$ .
The length $L$ in $x$ direction of this suspended D1-brane is
\be
L=2\int^\infty_{r_*}\f{dr}{r^2\s{B^2(r^2-\omega^2)(A^2r^2-1)-1}} \ .
\ee

We can find that $g(r)\simeq \mbox{const}-\f{1}{Ar}+\ddd$ in the $r\to\infty$ limit.
Thus the current flow is
\be
\f{dQ_R}{dt}=\f{T_{D1}}{A}-\f{T_{D1}}{A}=0 \ ,
\ee
and the energy flow is
\be
\f{dE}{dt}=\f{T_{D1}\omega}{A}-\f{T_{D1}\omega}{A}=0 \ .
\ee

The reason we find that $\f{dQ_R}{dt}=\f{dE}{dt}=0$
is that we employed the simple ansatz (\ref{xx})
and if we want to violate this we need to consider
truly time-dependent solutions which are
too complicated. The parameters which describe this system are
$\omega, A$ and $L$. Remember that in the
previous solution of thermal D1-branes, we had the constraint
(\ref{relationa}) from the regularity of the solutions and the
the R-charge current is determined from the total R-charge.
On the other hand, in the connected D1-brane solution, they are independent as
there is no thermal equilibrium even in the flavor sector.

Moreover, because $r_*>\omega$,
all of these suspended rotating D1-branes do not have any horizons in the
induced metric
\be
ds^2_{ind}=-(r^2-\omega^2)dt^2+2\omega g'dtdr+\left(\f{1}{r^2}+r^2x'^2+g'^2\right)dr^2 \ .
\ee

We can compute the time delay along the null geodesic in the induced metric
 similarly to the global AdS case.
The induced metric on the suspended solution can be rewritten as follows
\begin{align}
ds^{2}_{ind} &= -(r^{2} - \omega^{2}) \left( dt - \f{B \omega}{r^{2}\s{B^{2}(r^{2}
	- \omega^{2})(A^{2}r^{2} - 1)-1}}dr \right)^{2}\!\!\!\! +\!
\f{A^{2}B^{2}(r^{2} - \omega^{2})}{B^{2}(r^{2} - \omega^{2})(A^{2}r^{2} - 1) - 1}dr^{2} \no
	&= -(r^{2} - \omega^{2})dudv,
\end{align}
where
\begin{align}
	u& = t - \int dr \f{B(Ar^{2} + \omega)}{r^{2}\s{B^{2}(r^{2} - \omega^{2})(A^{2}r^{2} - 1)-1}} \ , \no
	v&= t + \int dr \f{B(Ar^{2} - \omega)}{r^{2}\s{B^{2}(r^{2} - \omega^{2})(A^{2}r^{2} - 1)-1}} \ .
\end{align}
Thus the time delay of massless scalar field propagating from a boundary $Q$ to another boundary $P$
is given by
\begin{align}
	\D t &= -\int_{\infty}^{r_{\ast}}dr \f{B(Ar^{2} - \omega)}{r^{2}\s{B^{2}(r^{2}
		- \omega^{2})(A^{2}r^{2} - 1)-1}} + \int^{\infty}_{r_{\ast}}dr \f{B(Ar^{2}
			 + \omega)}{r^{2}\s{B^{2}(r^{2}
		- \omega^{2})(A^{2}r^{2} - 1)-1}}\no
	&= \int_{r_{\ast}}^{\infty}dr \f{2AB}{\s{B^{2}(r^{2} - \omega^{2})(A^{2}r^{2} - 1)-1}} \ .
\end{align}
Actually,
we can show
 the fluctuations in $\theta$ direction around the D1-brane are described by
massless scalar fields only when $A=\f{1}{\omega}$.
We therefore find
\be
\Delta t=2\int^\infty_{r_*}\f{dr}{\s{(r^2-\omega^2)^2-\f{\omega^2}{B^2}}} \ ,
\ee
which is much larger than $L$ if $B\ll \f{1}{\omega}$. Thus, as opposed to the global AdS case, we
can have a large time delay $\Delta t\gg L$ and this is due to the strong interactions with
the deconfined $\CN=4$ SYM.

\section{Properties of Thermal Vacua from Rotating D-branes in AdS}
\hspace{5mm}
In this section we calculate various physical quantities
from our rotating D-branes and confirm that they are indeed
thermalized as we expected from the
induced metric.

\subsection{Action of Fluctuations}

First we need to show that the fluctuations on the rotating D-branes
see the horizon which we found from the induced metric.
Consider a D$p$-brane in the background metric (\ref{metgl}).

We choose the gauge where the worldvolume coordinates
$\xi^a = x^a,~~a=0...p$. The transverse coordinates are $x^I, I = p+1
\cdots 9$. Let us write the metric (\ref{fluc1}) as
\be
ds^2 = g_{ab} (x^a, x^I) dx^adx^b + G_{IJ}(x^a,x^K) dx^I dx^J \ .
\label{fluc1}
\ee
Expanding around a classical solution  $x_0^I(x^a)$,
\be
x^I(x^a) = x_0^I (x^a) + y^I(x^a) \ ,
\label{fluc3}
\ee
we find that upto terms of order $O(y^2)$ the induced metric is
\be
\gamma_{ab} = \gamma^{(0)}_{ab} + M_{ab} + N_{ab} \ ,
\label{fluc4}
\ee
where $\gamma^{(0)}_{ab}$ is the induced metric for $x^I_0$.
The matrices $M$ and $N$ are given by
\ba
M_{ab} & = & (\partial_I g_{ab})y^I + G_{IJ} (\partial_a y^I
\partial_b x_0^J + \partial_b y^J
\partial_a x_0^I) + (\partial_K G_{IJ})(\partial_a x_0^I \partial_b
x_0^J) y^K  \ ,\nn
N_{ab} & = & G_{IJ} \partial_a y^I \partial_b y^J +
(\partial_K G_{IJ})(\partial_ay^I\partial_bx_0^J + \partial_b y^J
\partial_a x_0^I)y^K + \nn
& & \frac{1}{2} (\partial_K\partial_L G_{IJ})y^Ky^L
\partial_ax_0^I\partial_bx_0^J + \frac{1}{2} (\partial_I\partial_J
g_{ab})y^Iy^J \ .
\label{fluc5}
\ea
The DBI action is then
\ba
S_{DBI} &\!\! =\!\! & -T_p \int d^{p+1}\xi~{\sqrt{{\rm det}~\gamma_{ab}}} \nn
&\!\! =\!\! & S_0 \!+\! \frac{T_p}{2}
\int d^{p+1}\xi \sqrt{-\gamma_0} \left[ \gamma_0^{ab}N_{ab} - \frac{1}{2}
  \gamma_0^{ab}M_{bc} \gamma_0^{cd} M_{da}\!+ \! \frac{1}{4}
  (\gamma_0^{ab}M_{ba})^2\! +\! O(y^3) \right],\!\
\label{fluc6}
\ea
where $S_0$ is the action of the background classical solution
$x_0^I$.

\subsection{Fluctuations around Rotating Strings}

In general the fluctuations are coupled with each other and have
complicated actions. For fluctuations around the rotating D1 or F1 solution
considered above the action simplifies drastically.
In the $(u,v)$ coordinates introduced in (\ref{uvdef}), the matrix
elements of $M$ and $N$ yield the following simple result
\be
M_{uu} =0~,~~~~~M_{vv} = 2 (\partial_v y^\varphi) (\partial_v \varphi_0)
~,~~~~~~~~~~M_{uv} = (\partial_u y^\varphi) (\partial_v \varphi_0) \ ,
\ee
and
\be
N_{ab} = G_{IJ}\partial_a y^I \partial_b y^J - (\partial_a \varphi_0
\partial_b \varphi_0) (y^\theta)^2 \ .
\label{fluc7}
\ee
Plugging this into (\ref{fluc6}) we see that in the first term
only the first term of (\ref{fluc7}) survives since
$\gamma_0^{ab}\partial_a\varphi_0 \partial_b \varphi_0 = 0$ while the
contributions from the second and third terms precisely cancel each
other, leading to the following action for quadratic fluctuations
\be
S_2 = \frac{T_{D1}}{2}\int d^{2}\xi \sqrt{-\gamma_0} \gamma_0^{ab} G_{IJ}
(\xi^a,x_0^I) \partial_a y^I \partial_b y^J \ .
\label{fluc8}
\ee
In particular, the fluctuations of $\varphi,\theta$ i.e. all fluctuations in $S^q$ directions
 are minimally coupled
massless scalars on the worldsheet, while the fluctuations of the
boundary gauge theory spatial directions $x^i, i
= 1\cdots 3$ have an additional factor of $r^2$ coming from the fact
that $G_{ij} = r^2 \delta_{ij}$ along these directions.

For a D1 or F1 rotating with a constant angular velocity, it is easy to see that
all fluctuations perceive a horizon at $r=\omega$ and its temperature is $T_H=\f{\omega}{2\pi}$.
For example, the action of the fluctuations in $x^i$ direction, denoted by $X$, is given by
\be
S=-T_{D1}\int dtdr\s{1-\f{r^2}{r^2-\omega^2}(\de_{\tau}X)^2+r^2(r^2-\omega^2)(X')^2} \ .
\ee

\subsection{Retarded Green's Functions from Rotating D1-branes}

As we discussed above, a rotating D1 or F1 corresponds to a source in
the $(0+1)$-dimensional defect CFT. In this subsection we will discuss
the retarded Green's function of dual operators to the worldvolume fields.

For the transverse scalars in the $S^5$ directions, this can be done using
the standard bulk-boundary relation \cite{GKP,W} in the presence of
horizon \cite{SS}, since these are minimally coupled fields on the
worldsheet.
Consider for example the worldsheet field $\vp$.
This is a massless scalar. Let us, however,
generalize the analysis and
consider a scalar field $\Phi$ with arbitrary mass $m$ which propagates
in the induced metric (\ref{indf}). They correspond to the infinitely many
stringy excitations of open strings.
The equation of motion is given by
 \be
 ((r^2-\vp'(v)^2)\Phi')'+2\dot{\Phi}'-m^2\Phi=0 \ .
 \ee
where prime denotes derivative with respect to $r$ and dot denotes
derivative with respect to $v$.
 In particular, we consider the constant angular velocity case, i.e.,
 $\vp'(v)=\omega$.  Its exact solutions with energy $\Omega$ look like
 \be
\Phi(v,r)=e^{-i\Omega v}
 \left[A\left(\f{r+\omega}{r-\omega}\right)^{-\f{i\Omega}
{2\omega}}P^{i\Omega/\omega}_\kappa(r/\omega)
   +B\left(\f{r+\omega}{r-\omega}\right)^{-\f{i\Omega}{2\omega}}
Q^{i\Omega/\omega}_\kappa(r/\omega)\right]
 \ ,
\ee
where $\kappa=\f{\s{1+4m^2}-1}{2}$; $P^a_b(x)$ and $Q^a_b(x)$
 are the first and second kind of the Legendre bi-function. By
 imposing the ingoing boundary condition at the horizon $r=\omega$, we
 simply find $B=0$. The retarded Green's function of the operator $O$
 dual to the scalar field can be computed from the ratio
 $\f{\beta}{\ap}$ when we expand the scalar field as $\Phi\sim \ap
 r^{\kappa}+\beta r^{-1-\kappa}$ in the boundary limit $r\to\infty$.
 Finally we thus find that it is given by the ratio of gamma functions
 as follows \be
\la
 O(\Omega)O(-\Omega)\lb_R=
\omega^{2\kappa+1}~\f{\Gamma(\kappa+1-i\Omega/\omega)}
 {\Gamma(-\kappa-i\Omega/\omega)} \ ,\label{cord}
\ee
up to some
 numerical factor independent of $\Omega$.  Indeed, for generic values
 of $m$, it has non-trivial poles only
 in the lower half plane and has the correct property of retarded
 Green's functions of thermal systems. Particularly
 for a massless scalar field, the dependence
 on $\omega$ drops out in (\ref{cord}) by chance and the thermal
 effect is not explicit. However, the same is true for the retarded
 Green's functions in massless free field theory at finite
 temperature and thus we can still conclude that the flavor sector dual to the
 our probe D-string is thermalized.
 For example, this issue is clear by comparing the
 time-ordered two point function (\ref{thcor}) with the retarded two
 point function (\ref{retcorh}) of a harmonic oscillator given in the
 Appendix A. This thermal property will also be shown more
 clearly in the Brownian motion analysis given in the next subsection.

The fluctuations in the $x^i$ direction have an additional factor of
$r^2$ in front of the Lagrangian, and leads to the following purely ingoing
solution which coincides with the one in \cite{Bra}
\be
X^i(t,r) = \frac{r+i\Omega}{r}\left( \frac{r-\omega}{r+\omega}
\right)^{i\frac{\Omega}{2\omega}}.
\ee
For large values of $r$, the leading term is a constant, but the
coefficient of the subleading $1/r$ term vanishes. In this case,
however, the retarded Green's function cannot be simply read off from
the ratio of the subleading to the leading term, since the equation of
motion is different, and the analysis of \cite{SS} has to be redone.

\subsection{Brownian Motion of Endpoints for Rotating D1-branes}

The thermal nature of the state produced by time dependence becomes
clear from a slightly different calculation - the fluctuation
of the end-point of the string.
In \cite{Bra} and \cite{Brb} it has been shown that the fluctuations
of a string suspended from the horizon of a AdS black brane ended at
a flavor D-brane near the boundary of AdS are dual to Brownian
motion of the corresponding quark in the hot $\CN = 4$ gauge
theory. In this case the bulk black brane metric induces a worldsheet
metric which has a horizon. The fluctuations then reflect Hawking
radiation from the worldsheet horizon. The results in \cite{Bra} and
\cite{Brb} are consistent with the fluctuation-dissipation theorem.

In the D-brane solutions considered above, the bulk metric has no
horizon. However due to the motion of the D-brane, the induced metric
on the worldvolume can develop a horizon. Since the fluctuations of
\cite{Bra,Brb} comes purely from properties of the induced metric it
is natural to expect that a similar phenomenon appears in our case.

We therefore consider a rotating D1 (or F1) brane ending on a flavor
D-brane situated close to the boundary at $r = r_c$. One example of
such a flavor D-brane is the rotating D7-brane which will be discussed
in the next section and in this case we need to require $\omega<r_c$.
As usual a finite $r_c$ means that the dual description of the string
is a finite mass monopole (or quark), $m = T_{D1} r_c$. Following
\cite{Bra} we now compute the {\em normal ordered} Green's functions
of the end-point of the string, $\langle 0,U|:[ y^I (t,r_c) y^J
(t^\prime , r_c) ]:|0, U\rangle $ in a Unruh vacuum state $|0,
U\rangle$ of the worldsheet theory. This quantity is equal to the mean
square fluctuation of the end-point with a trivial divergence due to
zero point energy subtracted.  We will consider the case of a constant
spin.

For constant spin, $\varphi_0 (v) = \omega v$ the induced metric is
\be
ds^2_{ind} = (\gamma_0)_{ab}d\xi^ad\xi^b = 2drdv - (f(r) - \omega^2) dv^2 = -(f(r) - \omega^2) d\bu dv \ ,
\ee
where
\be
\bu = v - 2r_\star \ ,~~~~~~~r_\star = \int \frac{dr}{f(r) - \omega^2} \ .
\ee
The coordinate $r_\star$ is the tortoise coordinate on the worldsheet,
where the boundary is at $r_\star = 0$ and the horizon is at $r_\star = -\infty$.
The UV cutoff corresponding to the D7-brane means that the range of $r_\star$ is now
$ -\infty < r_\star < r_{\star,c}$, and we need to impose
Neumann conditions on the fluctuation field $y^\varphi$ at $r_\star = r_{\star,c}$.

The $\varphi$ fluctuations behave as minimally coupled massless scalars in the induced metric
produced by the rotating string solution. It turns out that for these fluctuations the UV cutoff can
be removed and we can treat the problem in the full range $ -\infty < r_\star < 0$.
the mode expansion is given by
\be
y^\varphi (\bt,r_\star) = \int \frac{d\nu}{2\pi \sqrt{2\nu}}\cos (\nu r_\star) [ a_\nu e^{-i\nu \bt} + h.c.] \ ,
\ee
where
\be
\bt \equiv \frac{1}{2} (\bu + v) \ .
\ee
Note that on the boundary $\bt = t$.
The oscillators $a^\dagger_\nu$ create "Schwarzschild" particles
from the Schwarzschild vacuum $|0,S\rangle : a_\nu |0,S\rangle = 0$.

Consider now the case where the background is $AdS_5 \times S^5$ in the Poincare patch, i.e. $f(r) = r^2$.
In the Unruh vacuum the correlators of $a_\nu$ are thermal with a temperature $T = 1/\beta = \omega / 2\pi$, i.e.
\be
\langle0,U| a^\dagger_\nu a_{\nu^\prime} |0,U\rangle = \frac{\delta (\nu - \nu^\prime)}{e^{\beta \nu} - 1} \ .
\ee
Using this it is straightforward to calculate the mean square displacement in the $\varphi$ direction,
\ba
\langle(\Delta y^\varphi (t-t^\prime))^2\rangle & = & \langle 0,U|:[y^\varphi(t) - y^\varphi (t^\prime)]^2:|0,U\rangle  \ ,\no
& = &
\frac{2}{\pi}\int \frac{d\nu}{\nu}\frac{\sin^2(\nu(t-t^\prime)/2)}{e^{\beta\nu}-1}  \ ,\no
& = & \frac{1}{2\pi} \log \left( \frac{\sinh (\pi (t-t^\prime)/\beta)}{(\pi (t-t^\prime)/\beta)} \right) \ .
\ea
Therefore

\ba \langle (\Delta y^\varphi (t-t^\prime))^2\rangle & \sim &
\frac{\pi (t-t^\prime)^2}{12 \beta^2} \ ,~~~~~~~~~~~~~~~~~~~\pi (t-t^\prime)
\ll \beta \ ,\no \langle (\Delta y^\varphi (t-t^\prime))^2\rangle &
\sim & \frac{(t-t^\prime)}{2 \beta} -\frac{1}{2\pi}
\log [2\pi (t-t^\prime)/\beta ]
\ ,~~~~~\pi (t-t^\prime) \gg \beta \ , \ea
exactly as in Brownian motion. The transition from the short time
ballistic behavior to the long time diffusive behavior happens at a
time scale set by the inverse temperature, which is the only scale in
the problem.  The result is identical for the $y^\theta$ fluctuations,
as well as fluctuations in the $S^3$ directions.

The fluctuations in the $\vx$ directions are not conformally coupled scalars.
The equation is in fact identical to the equation for fluctuations of the end point of a string
in the background of a BTZ black hole, as in
\cite{Bra}. In this case, the UV cutoff at $r=r_c$ is essential.
The final result is \cite{Bra}
\ba
\langle [\Delta y^i (t-t^\prime)]^2 \rangle & \sim & \frac{(t-t^\prime)^2}{m\beta} \ ,~~~~t \ll m\beta^2  \ ,\no
& \sim & \beta |t-t^\prime| \ ,~~~~t \gg m\beta^2 \ .
\ea
Unlike fluctuations in the internal $SO(6)$ space, Brownian motion in
the physical space on which the gauge theory lives is present only if the mass of the monopole or quark
is finite. This is of course expected. Furthermore the time scale which controls the
 transition from a ballistic behavior is not simply the inverse temperature, but
involves the mass in an essential way.

As argued above a rotating D1-brane corresponds to a time-dependent coupling in the defect ($0+1$)-dimensional
CFT. Brownian motion is a manifestation of thermalization of this defect CFT, or equivalently the
hypermultiplet sector of the theory. This Brownian motion is a physical consequence of the Hawking radiation
on the world-sheet and this corresponds to the thermalization of the hypermultiplet sector (or defect sector).
Notice that this is not directly related to the Hawking radiation in the bulk AdS which comes from the
thermalization of the $\CN=4$ super Yang-Mills sector. Indeed their temperatures can be chosen differently.

\subsection{Conductivity from Rotating D3-branes}

For rotating D$p$-branes of section (\ref{Dpp}), there are several other response functions which carry
signatures of thermalization. A good example is a
probe D3-brane configuration obtained by taking T-dual of the rotating D1 in $x,y$ direction.
This is a non-supersymmetric configuration in $AdS_5\times S^5$.
The rotating D3-brane solution is obtained from (\ref{dpsolone}) and (\ref{dpsoltwo}) by setting $p=3$.
Its induced metric is given by the black hole geometry (\ref{genind}) at $p=3$
and its Hawking temperature is
$T_{H}=\f{\s{3}\omega}{2\pi}$.

The worldvolume can now contain a nontrivial gauge field, and the DBI action of the D3-brane looks like
\ba
{\cal L}&=&\s{r^4+r^6\vp'^2-r^2\dot{\vp}^2-(1+r^2\vp'^2)F_{tx}^2-r^4F_{tr}^2+(r^2-\dot{\vp}^2)r^2F_{xr}^2 + 2r^{2}F_{rx}F_{tx}\dot{\vp}\vp'  } \ ,\no
&=&\s{\f{r^8}{r^4+r^2\omega^2+\omega^4}-\f{r^4}{r^6-\omega^6}F_{\tau x}^2-r^4F_{\tau r}^2+r^2(r^2-\omega^2)F_{xr}^2} \ , \label{DBID3}
\ea
This shows that the gauge field is propagating in a space-time with a horizon.

We would like to calculate the electric conductivity in the probe D3-brane background by applying the method
developed in AdS black holes \cite{SS,Ha}. We regard the abelian gauge field on the probe D3-brane as an external
gauge field for which we define the AC conductivity $\sigma(\nu)$, where $\nu$ is the frequency. In other words,
here we are thinking the charge with respect to the global flavor symmetry as the electric charge
to define the conductivity.

The DBI action (\ref{DBID3}) can be expressed using the induced metric $g_{ind}$:
\begin{align}
	L_{D3} = \s{-\det g_{ind}}\s{1+ \f{F_{\ap\beta}F^{\ap\beta}}{2}} \ .
\end{align}
It follows that the equations of motion are
\begin{align}\label{D3EOM}
	\nabla_{\ap}\left( \f{F^{\ap\beta}}{\s{1+ F^{2}/2}}\right) = 0 \ .
\end{align}

Before computing the conductivity, we have to find a configuration of
gauge field with non zero gauge potential
$A_{a} = (A_{t}(r) = \Phi (r),0,0,0) $ which obeys the following equation:
\begin{align}
	\partial_{r}\left( \f{r^{2} \s{\f{r^{4}+r^{2}\omega^{2} + \omega^{4}}{r^{4}}} \Phi'}{
	\s{1- \f{r^{4}+r^{2}\omega^{2} + \omega^{4}}{r^{4}}}\Phi'^{2}} \right) = 0 \ .
\end{align}
We can easily find a solution with the boundary condition $\Phi(r)=0$ at $r=\omega$:
\begin{align}\label{GaugePot}
	\Phi (r) = \int_{\omega}^{r} ds\f{s^{2}}{\s{(1+ C s^{4})(s^{4} + s^{2}\omega^{2} + \omega^{4})}} \ .
\end{align}
The gauge potential behaves near $r\sim \infty$ like
\begin{align}
	\Phi (r) = \m - \f{\rho}{r} + \dots \ ,
\end{align}
and $(\m, \rho)$ stand for the chemical potential and the electric density, respectively.
Accordingly, we can compute the electric density using (\ref{GaugePot}) as follows:
\begin{align}
	\rho = \lim_{r\rightarrow \infty}\Phi'(r) r^{2} = C^{-1/2} \ .
\end{align}

\begin{figure}[htbp]
\begin{center}
	\includegraphics[width=7cm]{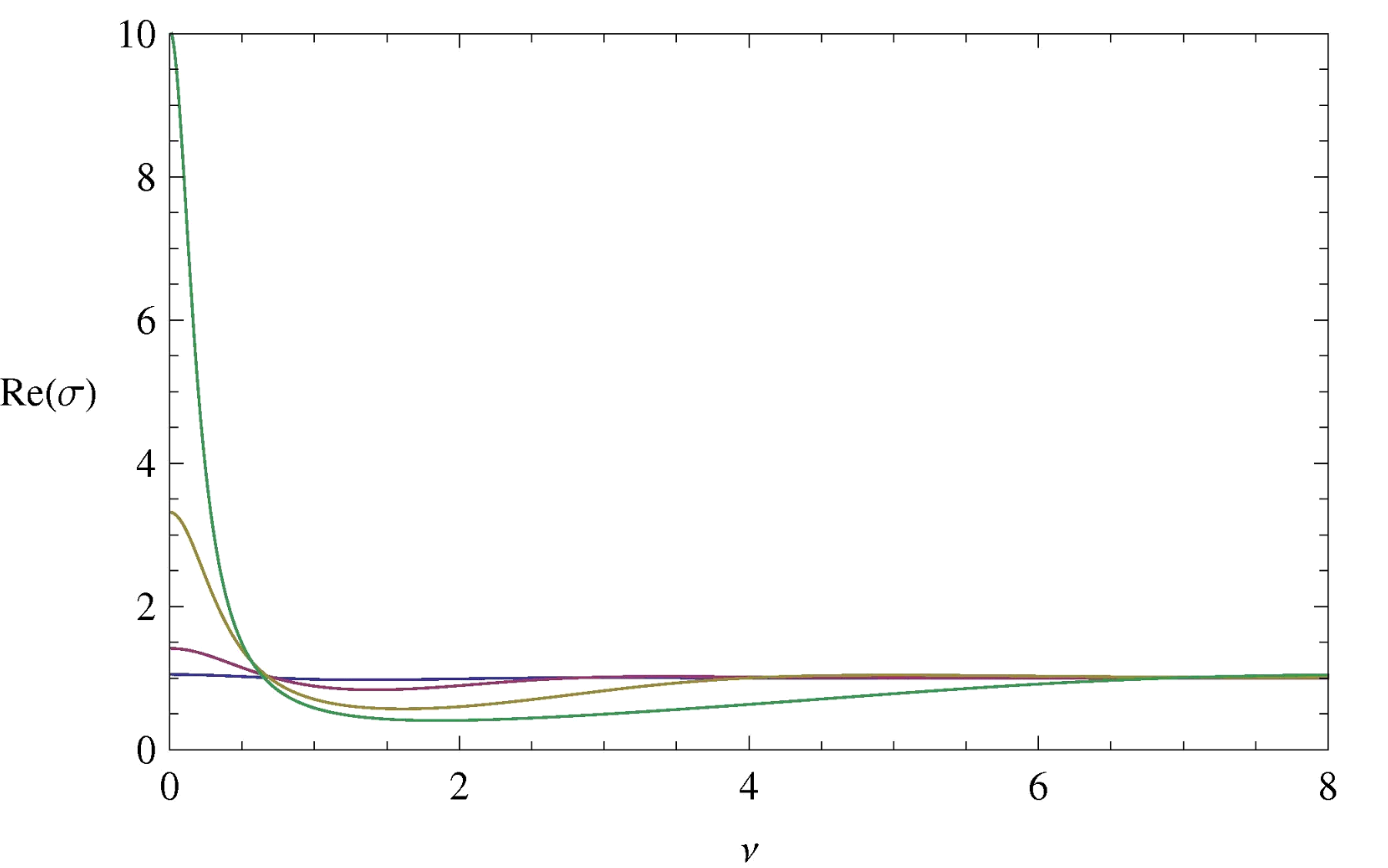}
		\hspace*{0.5cm}
	\includegraphics[width=7cm]{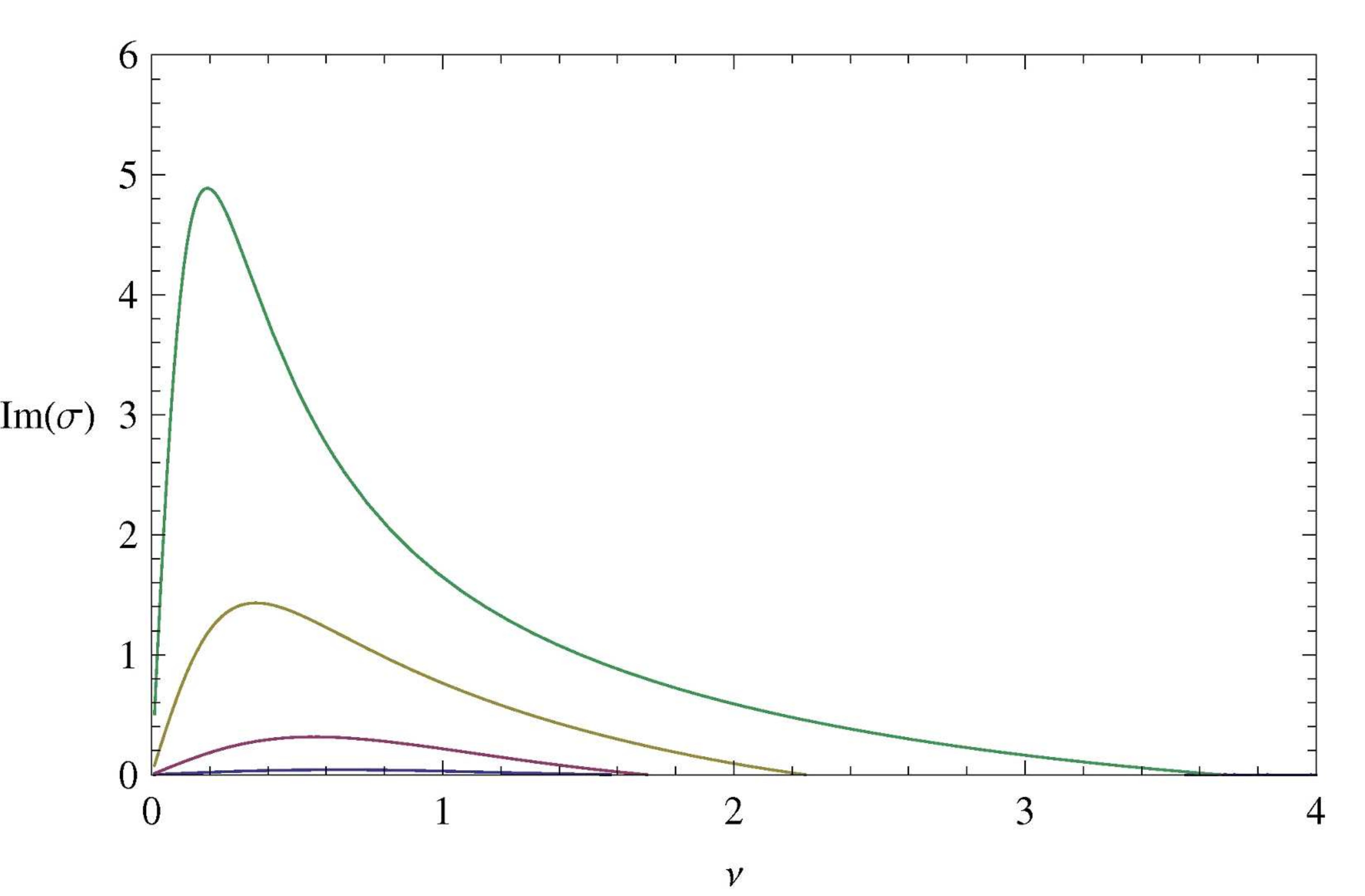}
\end{center}
\caption{The frequency dependence of the conductivity. Four lines are depicted for $\rho =
10, \s{10},1$ and $1/\s{10}$, respectively when $\omega =1$. The real part of the
conductivity approaches to Re\,$\sigma (\n) =1$ line as $\rho$ decreases, while the imaginary
part becomes zero.}
\label{Conductivity}
\end{figure}

Now we consider a small perturbation $A_{x}=A_{x}(r)e^{-i\n t}$ on this background.
We use $z\equiv 1/r$ coordinate for simple calculation in the following.
The equation we solve is that
\begin{align}\label{AxEOM}
	A_{x}(z)'' - \f{z^{3}(6z^{6}+z^{4}+z^{2}-2) + \tilde C z(z^{2}+z^{2}+4)}{(1-z^{6})(z^{4}+
	\tilde C)}A_{x}(z)' + \f{\tilde \n^{2} A_{x}}{(1-z^{2})(1-z^{6})} = 0 \ ,
\end{align}
where we rescaled $z\rightarrow z/\omega $, and define $\tilde C = C\omega^{4}$ and
 $\tilde\n = \n /\omega$.
The gauge field is an ingoing wave near the horizon $z=1$, and it thus can be expanded as
follows:
\begin{align}
	A_{x}(z) = (1-z)^{-i \f{\tilde \n}{2\s{3}}} (1 + a_{1}(1-z) + a_{2}(1-z)^{2} + \dots) \ ,
\end{align}
while it behaves at the boundary $z\sim 0$ as
\begin{align}
	A_{x}(z) = A_{x}^{(0)} + A_{x}^{(1)}z + \dots \ .
\end{align}
We can obtain the conductivity
\begin{align}
	\sigma (\n) = -\f{i A_{x}^{(1)}}{\n A_{x}^{(0)}} \ .
\end{align}
In Fig.\ \ref{Conductivity} we plot the conductivity obtained by solving (\ref{AxEOM}) numerically.
The behavior that Re\,$\sigma(\nu)$ approaches to a finite constant is known to common in
(2+1)-dimensional critical theories \cite{Ha}. As opposed to the purely bulk calculation of the conductivity
from the AdS charged black holes, where we have Re\,$\sigma(\nu)\propto \delta(\nu)$,
instead we observe a smooth Drude-like peak both in Re\,$\sigma(\nu)$ and Im\,$\sigma(\nu)$
near $\nu=0$ in Fig.\ \ref{Conductivity}.
This is because our probe approximation introduces dissipation into the bulk and has the
finite DC conductivity. A very similar behavior has been already obtained in \cite{KB,HPST} from the probe analysis
of D-branes in AdS black holes. Here we managed to get a similar physics by using a rotating D-branes in pure AdS.

\subsection{Comments on Entropy for Rotating D-branes}

Before we finish this section, we would like to comment on a possible definition of
 entropy of rotating D-branes.
Since we find a thermal horizon in the induced metric, it is natural to expect that there exists
non-trivial entropy in such a system in spite of the absence of real black holes.

There is one natural candidate of such an entropy
for probe D$p$-branes: it is calculated from their free energy by Wick rotating the
time coordinate $\tau$ in (\ref{genind}). Consider first a rotating D1-brane in $AdS_5\times S^5$.
The free energy is simply calculated from the DBI action as follows
\be
F_1=T_{D1} R^2\int^{\infty}_{\omega} dr \ ,
\ee
where we revive the AdS radius $R$ for convenience.
The entropy is obtained by taking derivative w.r.t. $T_H=\f{\omega}{2\pi}$
\be
S_1=-\f{\de F_1}{\de T_H}=2\pi R^2 T_{D1}=2\pi \s{\f{N}{\pi g_s}}=2\s{\pi}\f{N}{\s{\lambda}} \ .
\ee
The linear dependence of the rank $N$ of gauge group agrees with what we expect for the flavor
contribution in the planar limit.

Let us move on to the higher-dimensional D-branes. As one such example, we would like to consider the
rotating D3-brane in $AdS_5\times S^5$ discussed in the previous subsection.
Then its free energy looks like
\ba
F_3&=&T_{D3}R^4V_2\int^{r_{\infty}}_\omega dr\f{r^4}{\s{r^4+\omega^2r^2+\omega^4}} \ ,\no
&=&T_{D3}R^4V_2\left(\f{r_{\infty}^3}{3}-\f{1}{2}r_{\infty}\omega^2+k\omega^3
\right) \ ,
\ea
where $k\simeq 0.135$  is a numerical constant; $V_2$ is the area of non-compact space defined by the intersection
of the D3 with the AdS boundary. We also kept the UV cut off $r_{\infty}(\to \infty)$ explicitly.
By taking the derivative with respect to the temperature $T_H=\f{\s{3}\omega}{2\pi}$,
we get the following entropy
\be
S_3=-\f{\de F_{3}}{\de T_H}=\f{2\pi}{\s{3}}T_{D3}R^4L^2\left(r_{\infty}\omega-3b\omega^2\right)=\f{2NV_2}{3}
\left(\f{T_H}{a}-\ti{k}T^2_H\right),
\ee
where $a=1/r_{\infty}$ is the lattice spacing (or UV cut off) of the
dual CFT and $\ti{k}=2\pi\s{3}k\simeq 1.47$ is a numerical
constant. Thus this entropy is UV divergent and this behavior is
always true for rotating D$p$-branes with $p\geq 1$. One may think
this is unnatural because the entropy which we calculated in this way
should be the thermal entropy of the hypermultiplet, which should be
finite.  This might suggest that the above way of computation of
entropy via our naive Wick rotation is not correct.  However, we would
like to propose a possible interpretation of this divergent entropy.
Recall that in our probe calculation, the interactions with the $\CN=4$
SYM sector were already incorporated as the supergravity
background. This means that the $\CN=4$ SYM sector was already traced
out in the density matrix.  Therefore the entropy which we calculated
in the above should be regarded as the entanglement entropy between
the flavor sector and the $\CN=4$ SYM sector. In thermal equilibrium,
we can neglect the entanglement.  However, in our non-equilibrium
systems where the two different temperatures coexist, there may be
non-trivial entanglement entropy which can be UV divergent. Notice
that the UV divergence of the standard entanglement entropy is
holographically explained by the infinite volume of AdS space
\cite{RT} and the origin of the divergence here looks similar.  It is
a quite intriguing future problem to confirm this from field theory
calculations.

\section{Rotating D7-branes in $AdS_5 \times S^5$}
\hspace{5mm}

So far we have assumed that the rotating D$p$-brane is point-like in
$S^q$.  However, this class of D-branes misses important
supersymmetric configurations because the latter often wrap on $S^n\
(n\geq 1)$ in $S^q$. One such example is the probe D7-branes in
$AdS_5\times S^5$, which is dual to the $\CN=2$ flavor hypermultiplets
coupled to the $\CN=4$ super Yang-Mills \cite{KK,Myers,Erd}.  Notice that
in this case the space-time of the dual field theory is
($3+1$)-dimensional even for the probe D7-brane sector.  Motivated by
this, in this section we will study the rotating D7-branes in
$AdS_5\times S^5$. This system has two independent parameters i.e. the
distance $m$ between the D7-brane and the background D3-branes and the
angular velocity $\omega$. This setup has been already studied in
\cite{Bannon,Evans}. Though our D7-brane solution presented below is
essentially the same as the one in \cite{Bannon}, we will study the solution
from a different viewpoint focusing on the presence of the thermal horizon in the
induced metric.

\subsection{D7-brane World-Volume}

The D7-brane world-volume coordinates are extending in the $AdS_5$ and $S^3$ included in $S^5$
with a rotation in $\vp$ direction. This is specified by
\be
\theta=\theta(r)\ ,\ \ \ \vp=\omega t+g(r) \ . \label{timed}
\ee
The DBI action now becomes
\be
{\cal{L}}=-r^3\cos^3\theta\s{L}\ ,
\ee
where
\be
L=1+r^2\sin^2\theta\vp'^2-\f{\omega^2\sin^2\theta}{r^2}+r^2\theta'^2-\omega^2\sin^2\theta\theta'^2 \ .
\ee

The equation of motion for $\vp$ reads
\be
\f{\de}{\de r}\left(\f{r^5\cos^3\theta\sin^2\theta \vp'}{\s{L}}\right)=0 \ . \label{vpeom}
\ee
The equation of motion for $\theta$  is given by
\be
3r^3\cos^2\theta\sin\theta\s{L}-r^3\cos^4\theta\sin\theta\f{r^2 g'^2-\f{\omega^2}{r^2}-\omega^2\theta'^2}{\s{L}}
+\de_r\left[r^3\cos^3\theta\f{(r^2-\omega^2\sin^2\theta)\theta'}{\s{L}}\right]=0 \ . \label{theom}
\ee

The solutions to the equation (\ref{vpeom}) should be either
\be
g'(r)=0 \ , \label{solone}
\ee
or
\be
g'(r)=\f{1}{r^2\sin\theta}\s{\f{(r^2-\omega^2\sin^2\theta)(1+r^2\theta'^2)}{(A^2r^8\cos^6\theta\sin^2\theta-1)}} \ .
\label{soltwo}
\ee
where $A$ is an integration constant.

\begin{figure}[htbp]
\centering
\includegraphics[width=6cm]{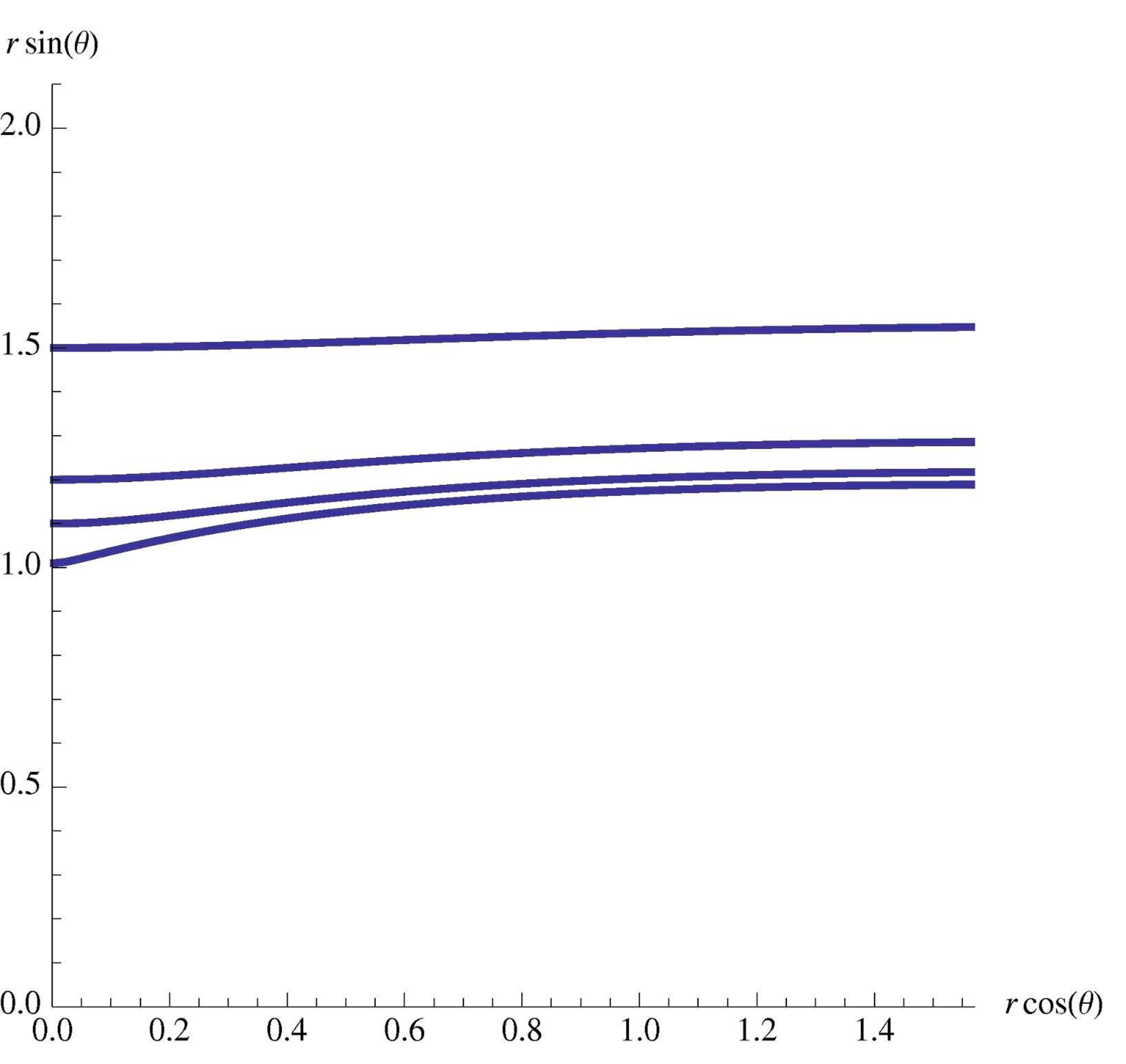}
\qquad
\includegraphics[width=6cm]{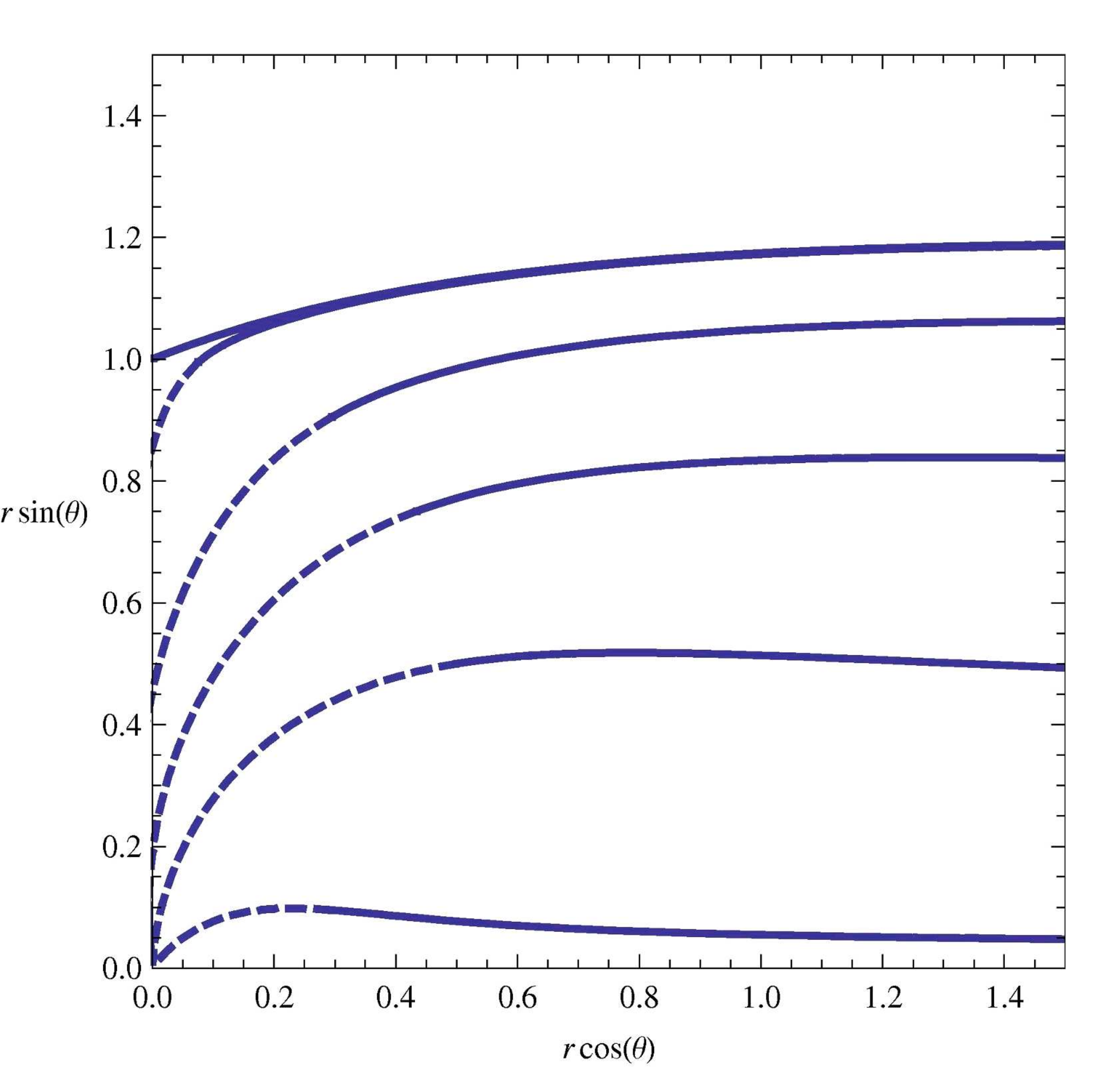}
\caption{The shapes of the rotating D7-brane with $\omega =1$. In the left
 panel where we consider $g'=0$ case, we depict the curves for
 $a=1.01,1.1,1.2$ and $1.5$.
In the right panel where we consider $g'\neq 0$, we depict them for $\ap
 =\pi/2, \pi/2.1, \pi/2.5, \pi/3, \pi/4, \pi/10$. The dotted regions are inside
 the horizon.}\label{d7shape}
\end{figure}

Let us first study the case (\ref{solone}). When $\omega=0$, we can
find the solution $r\sin\theta=m$.  This corresponds to the D7-brane
separated from D3-branes by the distance $m$, which is dual to the
mass of hypermultiplets.  Now consider solving the equation
(\ref{theom}) with a boundary condition $\theta = \pi/2$ at $r=a$
where $a$ is an arbitrary constant. The function $r(\theta)$ for
$\theta = \pi/2 + \delta$ for an infinitesimally small $\delta$ is
then determined by the equation of motion (\ref{theom}),
\be
r(\pi/2-\delta)\simeq a +\f{4a^3-3\omega^2
  a}{8(a^2-\omega^2)}\delta^2+\ddd \ .
\ee
Using this we can solve for
the function $r(\theta)$ numerically for $0 < \omega < a$.  In this
case the solutions does not differ qualitatively from the $\omega=0$
solution and there is no horizon, as can been from the left graphs in
the Fig.\ \ref{d7shape}.  If we increase $\omega$ such that
$\omega>a$, we cannot obtain smooth solutions from this ansatz
(\ref{solone}).

\begin{figure}[htbp]
\centering
\includegraphics[width=5cm]{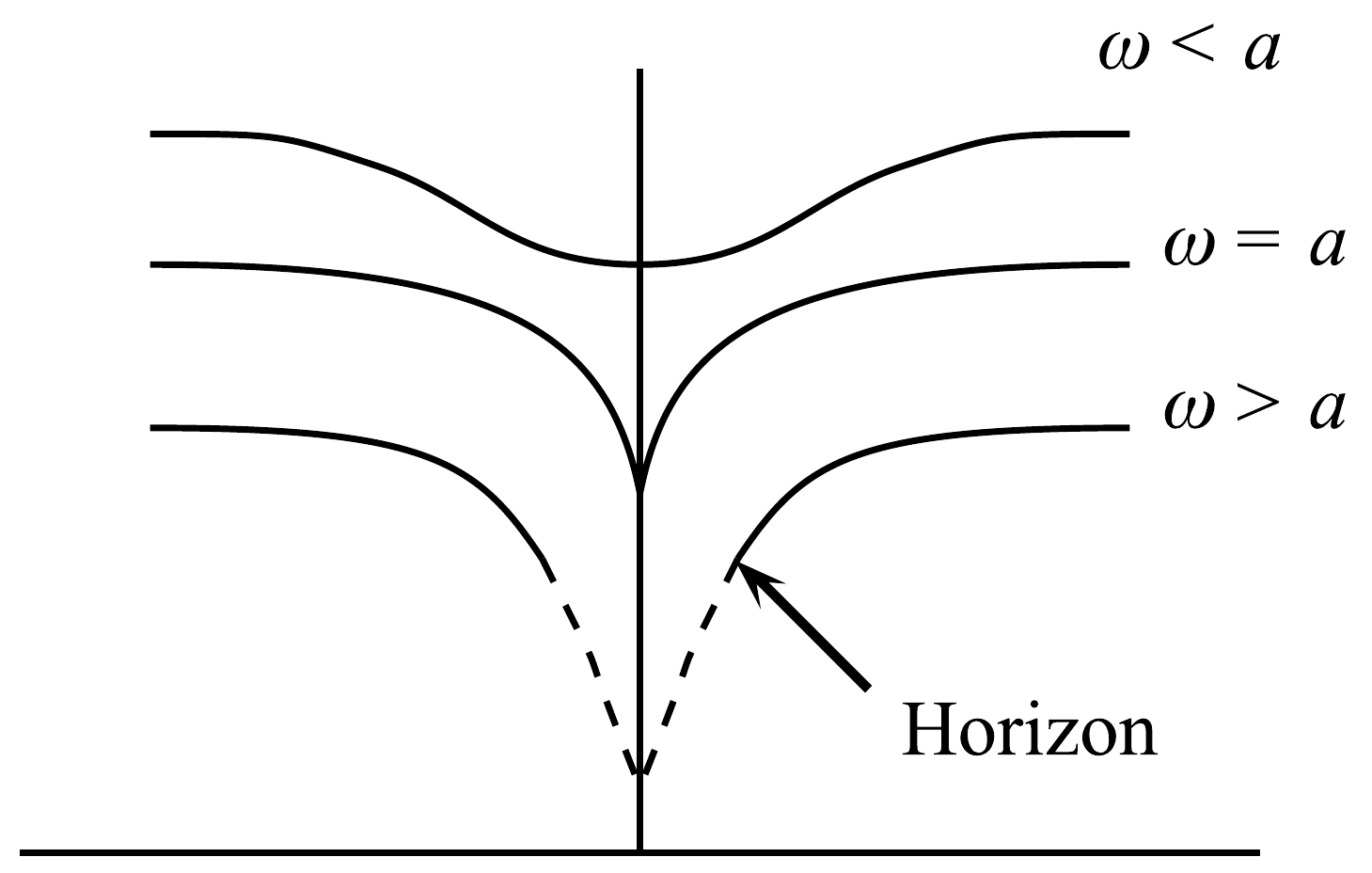}
\caption{The D7-brane is shaped like almost flat brane for $\omega < a$, and becomes singular when $\omega = a$. Then
it always have a horizon on its worldvolume for $\omega >a$.}
\label{fig:ShapeD7}
\end{figure}

Next we would like to consider the case when (\ref{soltwo}) is satisfied.
The denominator
$A^2r^8\cos^6\theta\sin^2\theta-1$ can become negative. To avoid the
resulting singularity in the
function $L$, we need to require that $r^2-\omega^2\sin^2\theta$ and $A^2r^8\cos^6\theta\sin^2\theta-1$ vanish
at the same time i.e.
\be
a^2=\omega^2\sin^2\ap\ , \ \ \ \ \ \ A^2a^8\cos^6\ap\sin^2\ap=1\ ,
\ee
where we set $r=a$ and $\theta=\ap$ at this point. This point will turn out to be the horizon.
If we expand the solution near this point, we find from the equation of motion (\ref{theom})
\be
\theta(r)\simeq \ap+\beta(r-a)+\ddd\ ,
\ee
where the (negative) constant $\beta(<0)$ is determined by
\begin{align}
3+ 2\beta\omega\cos\alpha - 3\beta^2\omega^2\sin^2\alpha =0 \ .
\end{align}
The solution to this equation for $\beta$ is
\begin{align}
\beta = \f{1}{3\omega\sin^2\alpha}(\cos\alpha - \s{\cos^2\alpha +
 9\sin^2\alpha}) \ .
\end{align}
Thus if we input the values $(a,\omega)$ we can uniquely find the solution numerically
by solving  (\ref{theom}) starting from the thermal horizon point $r=a$. They are shown in
the right graphs of Fig.\ \ref{d7shape}. Notice that the two solutions (\ref{solone}) and (\ref{soltwo})
of the D7-brane match at $\omega=a$ (or $\ap=\pi/2)$
as depicted in Fig.\ \ref{fig:ShapeD7}.

\vspace{1cm}

\subsection{Induced Metric and Thermal Horizon}

The induced metric reads
\ba
ds^2&=&-(r^2-\omega^2\sin^2\theta)dt^2+2\omega\sin^2\theta g'(r)dtdr+\cos^2\theta d\Omega_3^2 \no
&&\quad + \,\left(\f{1}{r^2}+\theta'^2
+\sin^2\theta g'^2\right)dr^2+r^2(dx^2+dy^2+dz^2) \ .
\ea
After the coordinate change $t=\tau+h(r)$ with
\be
h'(r)=\f{\omega\sin^2\theta g'}{r^2-\omega^2\sin^2\theta} \ ,
\ee
we can rewrite the metric as follows
\ba
ds^2&=&-(r^2-\omega^2\sin^2\theta)d\tau^2+\left(\f{1}{r^2}+\theta'^2
+\f{r^2\sin^2\theta g'^2}{r^2-\omega^2\sin^2\theta} \right)dr^2\no
&&\quad + \,r^2(dx^2+dy^2+dz^2)+\cos^2\theta d\Omega_3^2 \ .
\ea
Thus the horizon is identified with $r=\omega\sin\theta$ as already mentioned.
The topology
of horizon is given by $R^{3}\times S^3$.

\begin{figure}[htbp]
 \centering
\includegraphics[width=6cm]{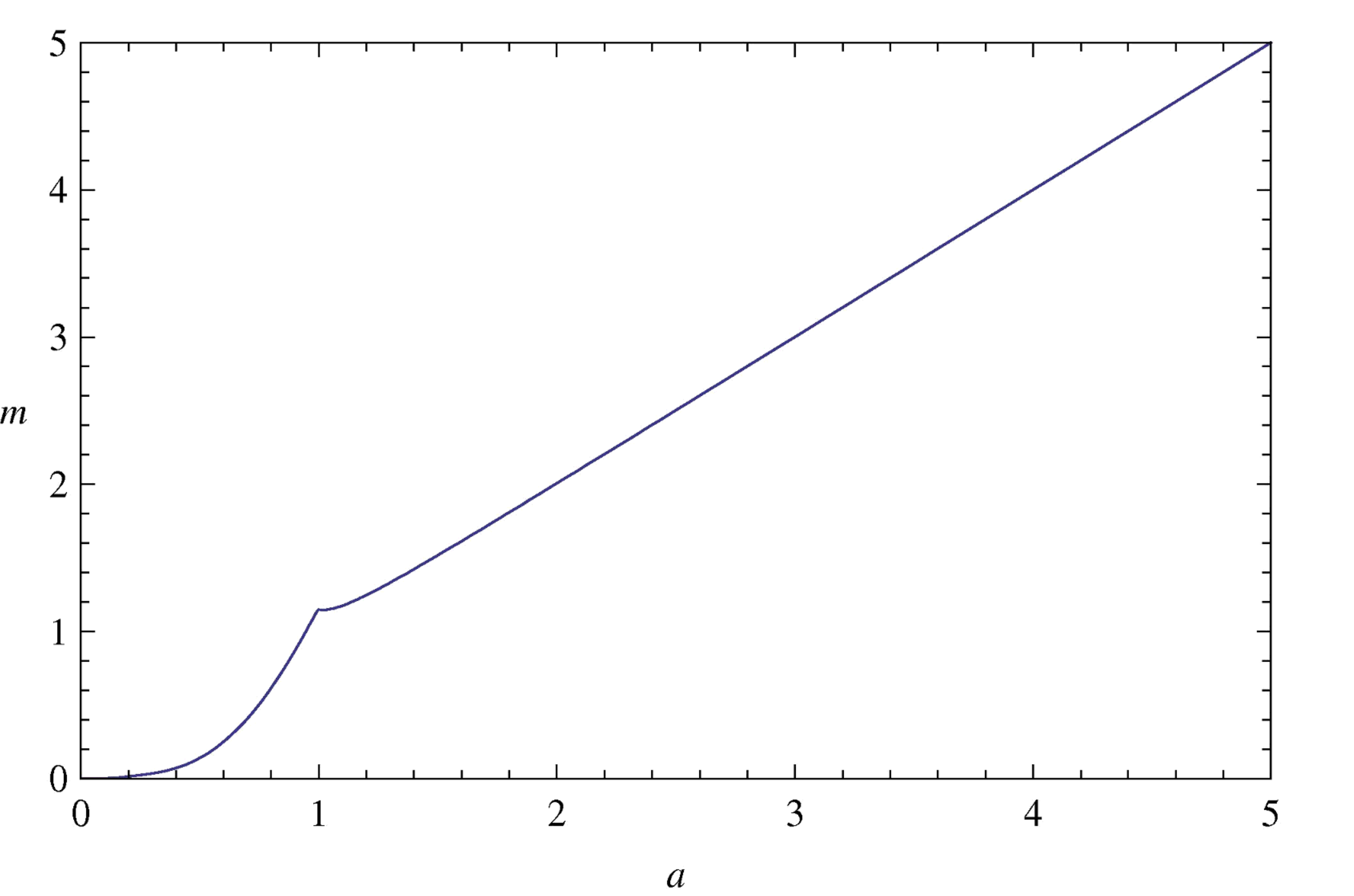}\hspace{1cm}
\includegraphics[width=6cm]{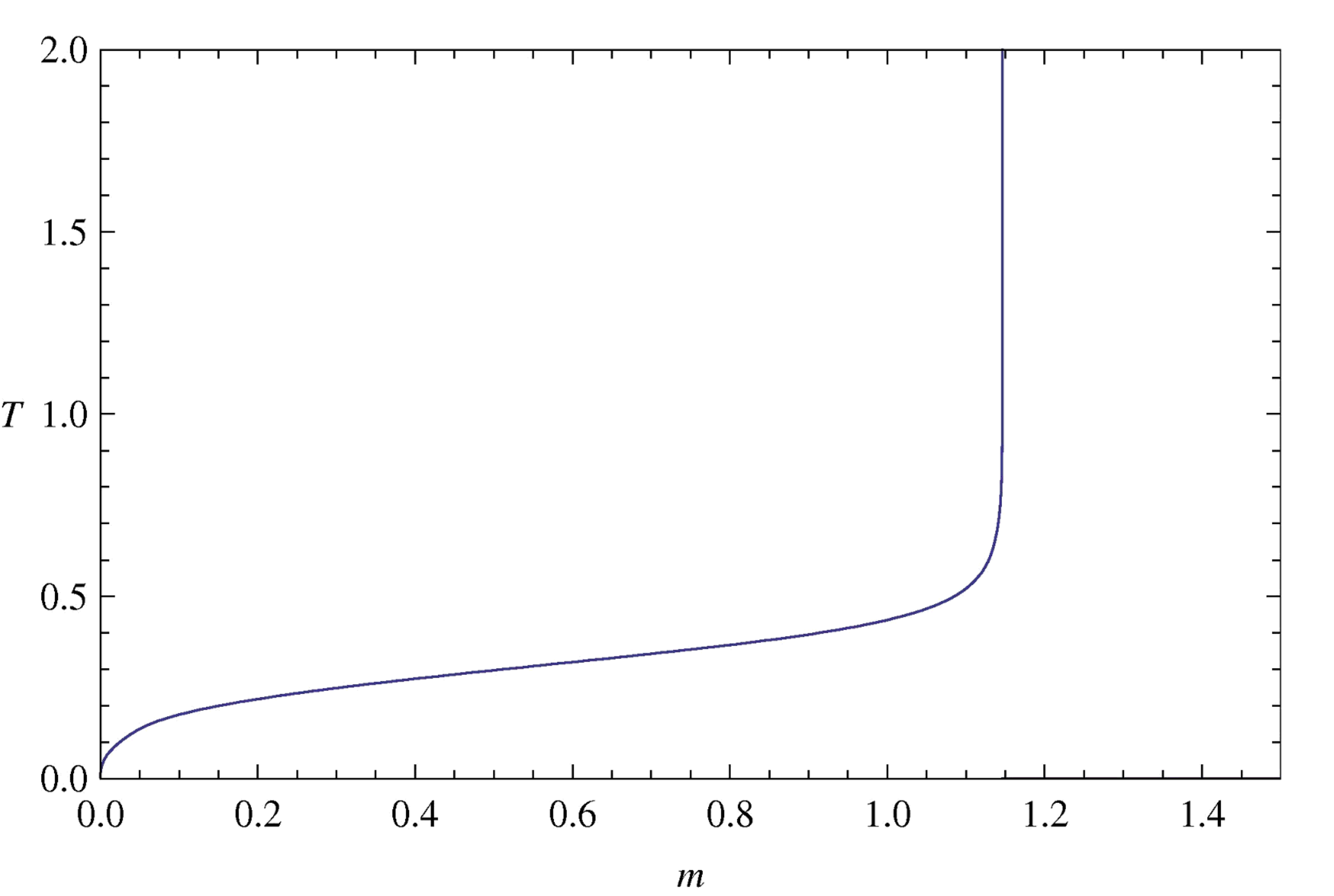}
\caption{[Left] The dependence of the function $m\equiv \lim_{r\rightarrow
 \infty} r\sin\theta$ on $a$ in the second solution (\ref{soltwo}).
 We set $\omega=1$ and connected the two solutions (\ref{solone}) and (\ref{soltwo}) at $a=1$.
 [Right] The dependence of the temperature on the mass $m$ at $\omega=1$. For $m\gtrsim 1.15$,
there is no horizon.}
 \label{d7mass}
\end{figure}

The Hawking temperature can be found from this metric
\begin{align}
T&= \f{\omega}{2\pi}\s{\f{\sin\a\tan\a(4\sin\a -
 \beta\omega(3+\beta\omega\cos
 3\alpha))}{1+\beta^2\omega^2\sin^2\alpha}} \ .
\end{align}
We can see that $T$ is monotonically increasing function of $a$ when we fix $\omega$. At $a=\omega$,
we find $T=\infty$, while at $a=0$, we have $T=0$. The flavor mass $m\equiv \lim_{r\to\infty}r\sin\theta$ of this solutions is plotted as a function of $a$ in the left panel of Fig.\ \ref{d7mass}. The dependence of the temperature on the mass is also presented in the right panel of Fig.\ \ref{d7mass}.

\subsection{Dual CFT Interpretation}

When the D7-brane does not rotate, our setup is the same as the standard D3-D7 system.
The dual CFT is described by the four-dimensional $\CN=2$ $SU(N)$ SQCD which is defined by
the $\CN=4$ SYM coupled to a $\CN=2$ flavor hypermultiplet in the large $N$ limit.
The time-dependence of $\vp(t)=\omega t$ (\ref{timed}) at the boundary $r=\infty$ is dual to the time-dependent
mass
\be
m(t)= me^{i\vp(t)}=m e^{i\omega t} \ .
\ee

In terms of the superpotential we have
\be
W=\mbox{Tr}[\Phi_1,\Phi_2]\Phi_3+\ti{Q}\Phi_3Q+m(t)\ti{Q} Q \ ,\label{sup}
\ee
where $\Phi_1,\Phi_2,\Phi_3$ are the transverse complex scalar fields on the $N$ D3-branes.
$(Q,\ti{Q})$ are the hypermultiplets from D3-D7 open strings.  We will also denote the
fermions in the hypermultiplet by $(\psi_{Q},\ti{\psi}_Q)$.  The original $SU(4)=SO(6)$
R-charge of $\CN=4$ super Yang-Mills theory is now broken to
\be
SO(4)\times U(1)_R=SU(2)_R\times SU(2)_L\times U(1)_R \ .
\ee
The $SO(4)$ rotates the $R^4$ described by $\Phi_1$ and $\Phi_2$. $U(1)_R$ corresponds to
the rotation of the remaining $R^2$ i.e. $\Phi_3$. The R-symmetry of $\CN=2$ super-Yang Mills
is given by $U(2)_R=SU(2)_R\times U(1)_R$. If we denote the spins under
$SU(2)_R\times SU(2)_L\times U(1)_R$ by $(j_R,j_L,q_{U(1)})$, the complex scalar fields $(Q,\ti{Q})$ in the hypermultiplet belongs to
$(1/2,0,0)$. On the other hand, fermions $(\psi_Q,\ti{\psi}_Q)$ belong to $(0,0,\pm 1)$. The transverse
scalar fields $\Phi_1,\Phi_2$ belong to $(1/2,1/2,0)$, while $\Phi_3$ to $(0,0,1)$.

Now we would like to come back to the superpotential (\ref{sup}). After integrating the
fermionic coordinates of the superfield, we find the time-dependent bosonic potential\footnote{
One may notice that if we redefine $\Phi_3$ as $\hat{\Phi}_3=\Phi_3 e^{i\vp(t)}$, then
the system is equivalent to the standard flavored $\CN=2$ SYM with
the  R-charge chemical potential $A_t=\omega$ only for the field $\hat{\Phi}_3$ without any time-dependence. However, from this viewpoint we need to take a time-dependent state and its time evolution will be described by this time-independent Hamiltonian. This is because we know from AdS/CFT
that the hypermultiplet is initially exited and has non-zero R-charge instead of the $\CN=4$ SYM sector.}
\be
\int dx^4\ \left[\bar{Q}\ov{(\Phi_3-me^{i\omega t})}(\Phi_3-me^{i\omega t})Q+
\ti{Q}(\Phi_3-me^{i\omega t})\ov{(\Phi_3-me^{i\omega t})}~\ov{\ti{Q}}\right]\ .
\ee
as well as the time-dependent fermion mass term
\be
\int dx^4\  me^{i\omega t}\ti{\psi}_Q\psi_Q+(h.c.)\ .
\ee
The AdS/CFT claims that the above time-dependent system will lead to a thermal vacuum.
These time-dependent potentials triggers a physics similar to that in
quantum quench and strong interactions
in the hypermultiplets will be expected to lead to a thermalization. However, since
the hypermultiplet is coupled to the $\CN=4$ SYM sector, the energy eventually will dissipate, as
we have explained in the simpler model of rotating D1-brane previously, schematically
summarized in Fig.\ \ref{fig:NonEqSystem}. It will be a quite interesting
future problem to confirm this prediction from the field theory calculations.

\section{Quantum Quench and Thermalization via AdS/CFT}
\hspace{5mm}
So far we have mainly studied stationary configurations of the
D$p$-branes. We would now like to
discuss truly time-dependent D-branes in AdS spaces and its interpretation via AdS/CFT correspondence.
The goal of this section is to holographically model quantum quench -
the time evolution of a system following a sudden
change of a parameter in a given quantum field theory, such as the mass of scalar fields.
The phenomenon of
quantum quench (see e.g. \cite{CCa,CCb,CCc,SCa,CCd,SCb} for approach from conformal field theories) has been intensively studied in condensed matter theory recently.

\subsection{Quantum Quench in Free Field Theory}

Some interesting aspects of the effect of quantum quench can be in fact observed in a
{\em free} quantum field theory \cite{CCc,CCd}.
Let us consider a $d$-dimensional massive bosonic free field theory with Hamiltonian
\begin{align}
	H = \int d^{d}k \left( \f{1}{2} \pi_{k}\pi_{-k} + \f{1}{2}\omega_{k}^{2}\phi_{k}\phi_{-k}\right) \ ,\label{freeh}
\end{align}
with a dispersion relation $\omega_{k}^{2} = k^{2} + m^{2}$.

Each momentum mode is independent of each other, and the propagator is
just that of a single harmonic oscillator.  We therefore consider a
single harmonic oscillator with momentum $k$ that initially lies in a
thermal state with temperature $\beta_{0}$, and then we quench its
frequency from $\omega_{0k}$ to $\omega_{k} $ by changing mass from
$m_{0}$ to $m$ at $t=0$.  At a later time $t$ after the quench, the
scalar field is given by $\phi_{k}(t) = \phi_{k}(0)\cos \omega_{k} t +
\pi_{k}(0) \sin \omega_{k} t /\omega_{k}$. The time ordered correlator
will be \cite{CCd}
\begin{align}\label{PropM}
	C_{\beta_{0}}(k; t_{1},t_{2}) &\equiv T\bigl< \phi_{k}(t_{1}) \phi_{-k}(t_{2}) \bigr> \no
	&= \f{1}{2\omega_{k}}e^{-i\omega_{k} |t_{1} - t_{2}|} + \left[ \f{\omega_{0k}}{4}\left(
	\f{1}{\omega_{0k}^{2}} + \f{1}{\omega_{k}^{2}}\right) \coth \f{\beta_{0}\omega_{0k}}{2}
	-\f{1}{2\omega_{k}} \right] \cos\omega_{k} (t_{1} - t_{2}) \no
	&\quad + \f{\omega_{0k}}{4}\left( \f{1}{\omega_{0k}^{2}} - \f{1}{\omega_{k}^{2}} \right)\coth\f{\beta_{0}\omega_{0k}}{2}\cos\omega_{k} (t_{1} + t_{2}) \ .
\end{align}
This result (\ref{PropM}) directly follows from the two point function (\ref{toh}) in a harmonic
oscillator as explained in the Appendix A of the present paper.

In real space,
the correlator is obtained by taking the Fourier transform of (\ref{PropM}).
The important fact is that
the third term in (\ref{PropM}), which violates time translation
invariance,
turns out to
decrease and eventually vanish after a long time under quite general conditions
\cite{CCc,CCd}. The time one has to wait to see this effect depends on
the wave-number $k$ and increases as $|k|$ decreases. Notice that the density matrix of
this time-dependent system is always pure since we start
from a pure state at zero temperature and it evolves in a unitary way under the
time-dependent Hamiltonian. Thus to find a thermal behavior we need a process of
coarse graining.

In this way, we can ignore this part and compare the others to the thermal propagator with
temperature $\beta$ (see (\ref{thcor}) in Appendix A for an elementary derivation)
\begin{align}
	G_{\beta}(k;t_{1},t_{2}) = \f{1}{2\omega_{k}}e^{-i\omega_{k} |t_{1} - t_{2}|}
	+ \f{\cos \omega_{k}(t_{1} - t_{2})}{\omega_{k} (e^{\beta \omega_{k}} - 1)} \ . \label{thpo}
\end{align}
We can read off the effective temperature by equating the coefficient of $\cos\omega_{k}(t_{1} - t_{2})$
\begin{align}
	\beta_{eff}(k) = \f{1}{\omega_{k}}\log \f{(\omega_{k} - \omega_{0k})^{2} +
	e^{\beta_{0}\omega_{0k}}(\omega_{k} + \omega_{0k})^{2} }{(\omega_{k} + \omega_{0k})^{2} +
	e^{\beta_{0}\omega_{0k}}(\omega_{k} - \omega_{0k})^{2}} \ . \label{tpfree}
\end{align}
It follows from this result that the effective temperature depends on each momentum mode.

It is clear that an effective temperature is obtained even when the initial state is the vacuum state
rather than a thermal state. In this case
\be
\beta_{eff,0}(k) = \frac{2}{\omega_k}\log \frac{\omega_k + \omega_{0k}}{|\omega_k - \omega_{0k}|} \ .
\ee

It is easy to extend the above calculation to a situation where the mass starts
from a value $m_0$ at $t=0$, changes to a value $m \neq m_0$ for $0 <
t < T$ and then finally changes back to $m = m_0$ for $t >
T$. Interestingly the late time correlation functions for $t \gg T$
are again thermal with a $k$-dependent effective temperature. This is
consistent with the fact that once the system thermalizes, it remains
thermal.

\subsection{Exact Toy Model from D1-branes}

The framework considered in this paper can be used to construct a very simple
toy model for a quantum quench in a strongly interacting theory.
This is provided by the
exact time-dependent solution of D1-brane
whose induced metric is given by the AdS Vaidya metric studied generally in (\ref{indf}).
In our current example, its two-dimensional induced metric looks like
\be
ds^2=2drdv-(r^2-\vp'(v)^2)dv^2.
\ee

To describe a quantum quench we can for example take ($v\equiv t-1/r$)
\be
\vp(v)=\vp_0(1+\tanh kv)\ ,
\label{phiform}
\ee
which approaches a step function in the large $k$ limit. Then the apparent horizon is situated at
\be
r=\vp'(v)=\f{k\vp_0}{\cosh^2 kv}\ .
\ee
In this model, the thermalization is occurred due to the time-dependent coupling (\ref{intti})
as we discussed before.

When the apparent horizon is slowly evolving, it is possible to define
a local temperature by considering the analytic continuation of outgoing modes
across the apparent horizon \cite{damour}. In our case
this leads to a ``local'' temperature given by $\vp^\prime(v)$. While
this expression is not strictly valid when the evolution is fast, we
can still use this to get a rough estimate of the temperature. In an
eikonal approximation, the
Hawking particles emitted at some point $(r,t)$ on the apparent horizon will
travel to the boundary along a line of constant $u = t + 1/r$ and will
therefore reach the boundary at time $t = u -1/r = v + 2/r$. Therefore
the temperature perceived at the boundary at time $t$ is given by
\be T_H(t)=\vp'(v)\ ,\ \ \ \ t=v+\f{2}{\vp'(v)}\ .
\label{timedelay}
\ee
where we need to first express $v(t)$ using the second equation of
(\ref{timedelay}) and then substitute this in the first equation to
obtain $T_H(t)$.

\begin{figure}
\begin{center}
\includegraphics[height=6cm,clip]{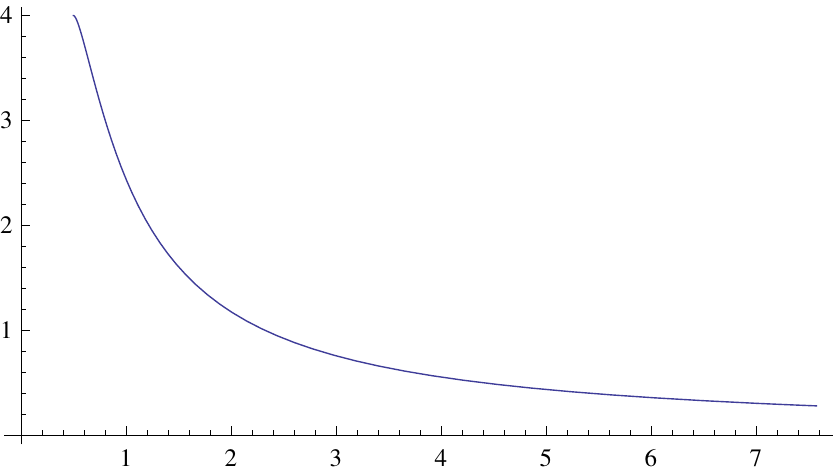}
\end{center}
\caption{The time-dependence of the effective temperature $T_H(t)$
in the D1-brane model under an eikonal approximation. We set $k=4$ and $\vp_0=1$.
}\label{efftem}
\end{figure}

For a $\vp(v)$ of the form (\ref{phiform})
this is plotted in Fig.\ \ref{efftem}.
The temperature is highest at $t=\vp'(0)=\vp_0$ and
decreases to zero at $t=\infty$.
Notice that the late time behavior is given by
\be
T_H(t)\sim \f{2}{t}\ .
\ee
This is a consequence of the conformal invariance of our D3-D1 system.

Quantum quench calculations in the literature have often been done in a
closed environment.  In such a case, a system got excited suddenly at
some time by an external force and will reach an equilibrium which are
static. On the other hand, in our system we find that the effective
temperature decreases to zero finally. This is because the
hypermultiplet sector, which are initially excited, is strongly
coupled to the $\CN=4$ SYM sector and thus the energy will dissipate.

\subsection{D3-D5 System with Time-dependent Mass}

Here we would like to turn to a more interesting example:
a holographic dual of quantum quench induced by time-dependent masses
of scalar and spinor fields. For this purpose we consider a probe D5-brane wrapped on
$AdS_4\times S^2$ in $AdS_5\times S^5$. This setup is supersymmetric and was first considered in \cite{KR}. It is dual to the four-dimensional $\CN=4$ SYM coupled to a three-dimensional defect CFT.
 If we write the metric of $AdS_5\times S^5$ as
\be
ds^2=r^2(-dt^2+\sum_{i=1}^3dx_i^2)+\f{dr^2}{r^2}+d\theta^2+\cos^2\theta d\Omega_2^2+\sin^2\theta d\ti{\Omega}_2^2 \ ,
\ee
then the D5 is wrapped on $\Omega_2$ and $(t,x_1,x_2,r)$ directions, while
it is trivially localized in $\ti{\Omega}_2$ and $x_3$ direction.
However, in actual numerical calculations, it is useful to perform
the following coordinate transformation
\be
\eta=r\sin\theta\ ,\ \ \ \  z=\f{1}{r\cos\theta}\ ,
\ee
for which the $AdS_5\times S^5$ metric looks like
\be
ds^2=(\eta^2+z^{-2})(-dt^2+\sum_{i=1}^3dx_i^2)+\f{d\eta^2}{\eta^2+z^{-2}}+\f{dz^2}{z^2+z^4\eta^2}
+\f{d\Omega_2^2}{1+z^2\eta^2}+\f{\eta^2d\ti{\Omega}_2^2}{\eta^2+z^{-2}}\ .
\ee
In this new coordinate, we can specify the profile of the D5-brane by the function
$\eta=\eta(t,z)$. Its DBI action reads
\be
S=-T_{D5}V_4\int dtdz \f{1}{z^4}\s{1+z^4\eta'^2-\f{\dot{\eta}^2}{(\eta^2+z^{-2})^2}}\ ,\label{QDBI}
\ee
where $V_4$ is the (infinite) volume of the $S^2\times R^2$. Its equation of motion looks like
\be
\f{\de}{\de z}\left(\f{\eta'}{\s{K}}\right)-\f{\de}{\de t }\left(\f{\dot{\eta}}{(1+z^2\eta^2)^2\s{K}}\right)-\f{2\eta\dot{\eta}^2}{z^4(z^{-2}+\eta^2)^3\s{K}}=0\ ,
\label{QEOM}
\ee
where $K\equiv 1+z^4\eta'^2-\f{\dot{\eta}^2}{(\eta^2+z^{-2})^2}$.

When $\eta$ can be treated as a small perturbation so that it satisfies
\be
z^2\eta'\ll 1\ ,\ \ \  \mbox{and}\ \ \  z^2\dot{\eta}\ll 1\ ,   \label{condqu}
\ee
we can approximate the equation of motion (\ref{QEOM})
by the simple wave equation i.e. $\eta''-\ddot{\eta}=0$. Therefore, if we require
that the defect sector in the dual CFT has the time-dependent mass $m(t)$, we can find the simple
solution under this approximation
\be
\eta(t,z)=m(t-z)\ .\label{appsolq}
\ee
However, in the interested regions where horizons appear, the condition (\ref{condqu}) is inevitably
violated. This is clear from the expression of $g_{tt}$ component in the induced metric
\be
g_{tt}|_{ind}=-(\eta^2+z^{-2})+\f{\dot{\eta}^2}{\eta^2+z^{-2}}\ ,
\ee
which should change its sign at apparent horizons. Since
in our particular setup, the area of a codimension two surface for fixed values of $t$ and $z$
is always only a function of $z$ (proportional to $z^{-2}$), the apparent horizon for our induced metric,
which is defined by the vanishing area change along null geodesics, actually
coincides with the points where $g_{tt}$ vanishes.

Thus we numerically integrate the equation of motion (\ref{QEOM}) under the two
boundary conditions presented below.
The first boundary condition is imposed at the AdS boundary with a UV cut off $z=z_0\ll 1$:
\be
\eta(t,z_0)=m(t-z_0)\ ,\ \ \ \ \ \f{\de\eta(t,z)}{\de z}\Bigr|_{z=z_0}=-\dot{m}(t-z_0)\ . \label{condqo}
\ee
The second one is at a initial time $t=t_0$:
\be
\eta(t_0,z)=m(t_0-z)\ ,\ \ \ \ \ \f{\de\eta(t,z)}{\de t}\Bigr|_{t=t_0}=\dot{m}(t_0-z)\ . \label{condqt}
\ee
Below we would like to analyze a specific choice of the mass function $m(t)$ defined by
\be
m(t)=m_0(1+\tanh kt)\ , \label{massti}
\ee
which describes a sudden quench for large $k$.

In both of the above conditions we employed the approximated solution (\ref{appsolq}).
For the first one (\ref{condqo}) this is clearly justified because $z_0$ is infinitesimally small.
On the other hand, for the next one (\ref{condqt}), the condition (\ref{condqu})
requires $t_0\ll \f{1}{\s{m_0k}}$ in our particular case (\ref{massti}).

\begin{figure}
\begin{center}
\includegraphics[height=6cm,clip]{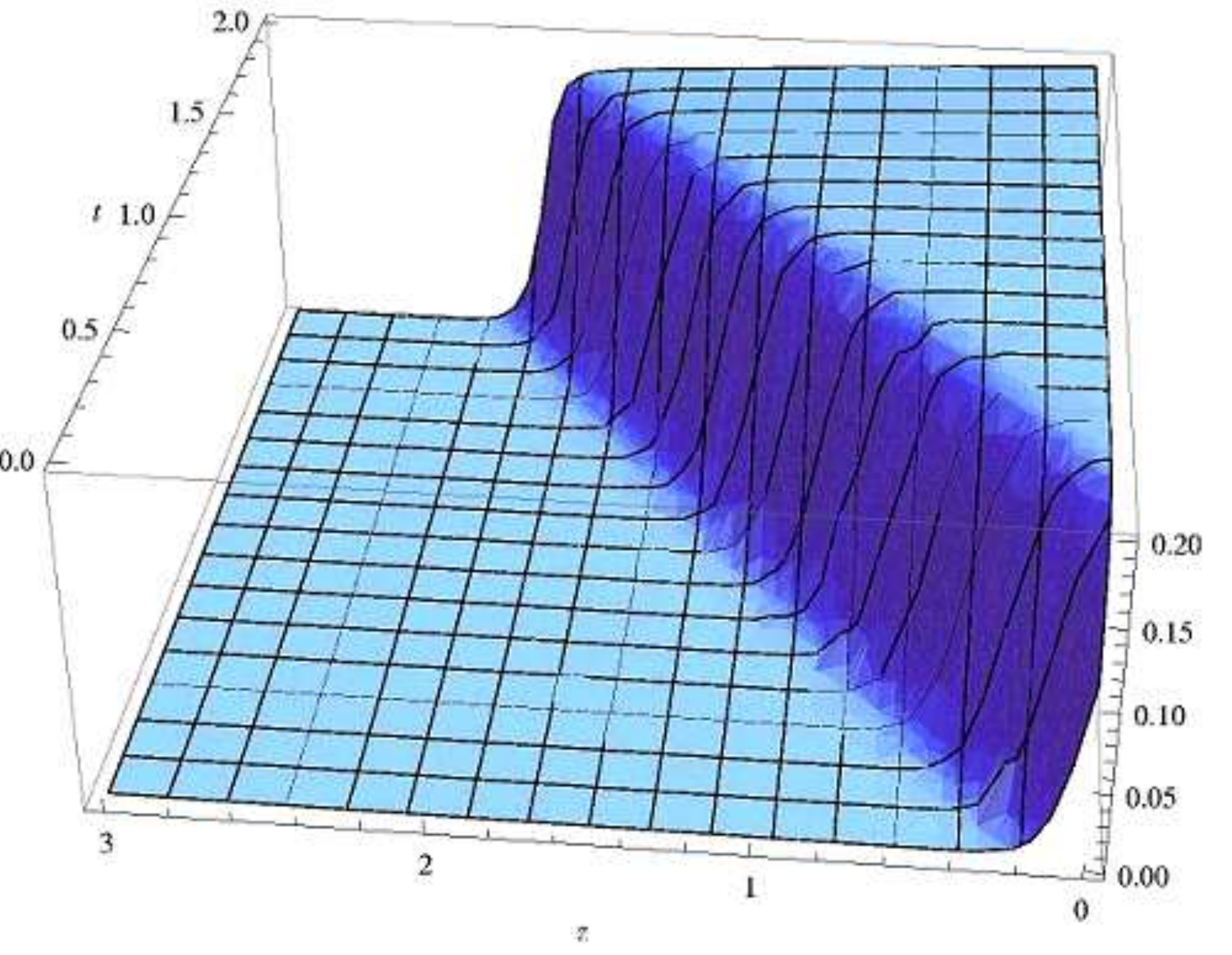}
\end{center}
\caption{The function $\eta(t,z)$ for the choice $k=10, m_0=0.1, t_0=0.01$ and $z_0=0.001$.
We plotted the region defined by $0.01<t<2$ and $0.001<z<3$.}\label{figone}
\end{figure}

\begin{figure}
 \begin{center}
 \includegraphics[height=6cm,clip]{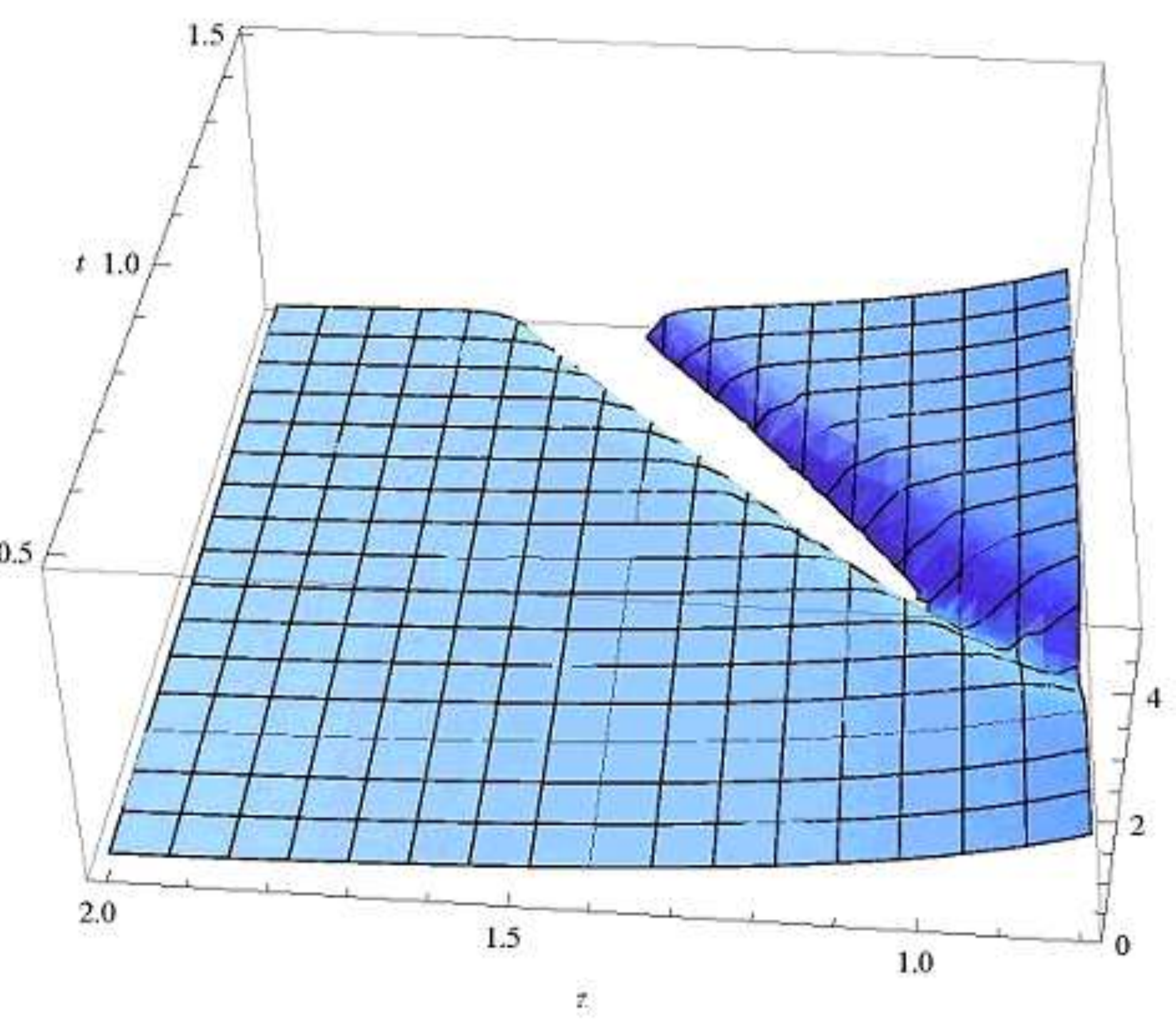}
 \end{center}
 \caption{The plot of $-g_{tt}(t,z)$ of the induced metric for the same choice as the previous
 plot. We concentrated on the region defined by $0.5<t<1.5$ and $0.8<z<2$. We plotted the graph only
 when $-g_{tt}>0$ and thus the hairpin-shape boundary in the middle represents the positions of the apparent horizons.}\label{figtwo}
 \end{figure}

Following this strategy, we plotted the function $\eta(t,z)$ in the
Fig.\ \ref{figone} for the particular choice of the parameters $k=10,
m_0=0.1, t_0=0.01$ and $z_0=0.001$.  As is obvious from this graph,
the kink, generated by the sudden mass change (\ref{massti}) in the
UV, propagates at the speed of light into the IR region. This is
qualitatively similar to our previous exact D1-brane solutions
discussed in the previous subsection. To show the existence of the
apparent horizon, we need to examine the sign of $g_{tt}$ of the
induced metric.  We plotted $-g_{tt}$ in Fig.\ \ref{figtwo} only when
it is positive. We can confirm a hairpin-shape boundary near the line
$t=z$, where $g_{tt}$ vanishes. Therefore we can conclude that our
time-dependent mass at the boundary generates a thermal horizon after
a certain time (estimates as $\sim \f{1}{\s{m_0k}}$). Its temperature
from an observer at the boundary decreases as time passes and
eventually goes to zero due to the dissipation of energy into the
bulk. From the dual CFT viewpoint, this clearly describes the process
of the thermalization in the open system when we change the mass of
scalar fields and fermions at some time.

Finally it is instructive to compare our result with that of free field theories.
In the free scalar field theory, after a sudden change of scalar mass, the system should
reach a sort of thermal equilibrium state. However, as we have reviewed briefly,
different momentum modes have different temperatures (\ref{tpfree}) \cite{CCc,CCd}.
This momentum dependent temperature
arises because the different momentum modes are not mixed with each other in the free field theory.
A similar issue has been known to occur in integrable systems \cite{Inte,CCd}, where the final equilibrium state
is described by the generalized Gibbs ensemble, due to the infinite number of conserved
quantities. However, our system is a strongly coupled
quantum field theory as we work in the supergravity regime and we expect substantial mixings between states
with different momenta. Indeed, our holographic analysis shows that the effective temperatures
for all momentum modes in the dual CFT are originated from the single apparent horizon at a fixed time.
Therefore our holographic description predicts that the final temperatures for each momenta
at the equilibrium should be the same in strongly coupled field theories.

\vskip6mm
\noindent
{\bf Acknowledgments}
S.R.D. would like to thank Ganpathy Murthy for introducing him to the
area of quantum quench and for many discussions. T.T. would like to thank
Shinsei Ryu for valuable discussions on non-equilibrium phenomena
in condensed matter physics and helpful comments on the draft of this paper.
  It is a great pleasure
to thank M. Fujita, S. Hellermann, W. Li, H. Liu, O. Lunin, G. Mandal,
S. Mathur,
R. Myers,
J. Ohkubo and Y. Srivastava for useful discussions.
S.R.D. would like to thank IPMU for hospitality and for
providing a superb intellectual atmosphere, and
Perimeter Institute for hospitality during the completion of the paper.
T.N. would like to thank IPMU, NTU, NTNU and NCTS for hospitality during
the completion of the paper.
T.T. is grateful to the
hospitality of the Center for Theoretical Physics at MIT and the Erwin
Schrodinger International Institute for Mathematical Physics in Vienna
where some parts of this work have been done.
The work
of S.R.D. is partially supported by a NSF grant NSF-PHY-0855614. The
work of T.N. is supported by JSPS Grant-in-Aid for Scientific Research
No.22$\cdot$2839. The work of T.T. is also supported in part by JSPS
Grant-in-Aid for Scientific Research No.20740132, and by JSPS
Grant-in-Aid for Creative Scientific Research No.\,19GS0219.  T.T. is
supported by World Premier International Research Center Initiative
(WPI Initiative), MEXT, Japan.

\newpage
\appendix
\section{Thermal Correlation Functions and Mass Quench in Free Theories}

\subsection{Thermal Correlation Functions}

Here we would like to give an elementary derivation of thermal correlation functions for a harmonic
oscillator and a free scalar field theory. We start with the standard Hamiltonian of a harmonic oscillator
\be
H=\f{1}{2}\hat{p}^2+\f{\omega^2}{2}\hat{x}^2=\omega\left(a^\dagger a+\f{1}{2}\right) \ ,\label{harmo}
\ee
where
\be
a^\dagger=\f{1}{\s{2}}\left(\s{\omega}\hat{x}-i\f{\hat{p}}{\s{\omega}}\right) \ ,
\ \ \ \ a=\f{1}{\s{2}}\left(\s{\omega}\hat{x}+i\f{\hat{p}}{\s{\omega}}\right) \ .
\ee
The vacuum $|0\lb$ is define by $a|0\lb=0$ and we can define the normalized number
state $|n\lb=\f{1}{\s{n!}}(a^\dagger)^n|0\lb$ as usual.

Using the commutation relation $[a,a^\dagger]=1$, we obtain
\be
\hat{x}(t)=\f{1}{\s{2\omega}}\left(ae^{-i\omega t}+a^\dagger e^{i\omega t}\right) \ , \label{evol}
\ee

The finite temperature density matrix for the inverse temperature $\beta$ is give by
\be
\rho=e^{-\beta H}=\f{\sum_{n=0}^\infty e^{-\left(n+\f{1}{2}\right)\beta\omega}|n\lb \la n|}{Z_0} \ , \ \ \ \
\ \ \ \left(Z_0\equiv \f{1}{2\sinh\f{\beta\omega}{2}}\right) \ .
\ee

We can calculate various two point functions as follows. First of all, we obtain the Wightman function
\be
\la \hat{x}(t)\hat{x}(0)\lb \equiv \mbox{Tr}[e^{-\beta H}  \hat{x}(t)\hat{x}(0)]
=\f{\cosh\left(\f{\beta\omega}{2}-i\omega t\right)}{2\omega \sinh\f{\beta\omega}{2}} \ .
\ee
Next, the time ordered two point function is given by
\be
T\la \hat{x}(t)\hat{x}(0)\lb
=\f{\cosh\left(\f{\beta\omega}{2}-i\omega |t|\right)}{2\omega \sinh\f{\beta\omega}{2}}
=\f{1}{2\omega}e^{-i\omega|t|}+\f{\cos\omega t}{\omega (e^{\beta\omega}-1)} \ . \label{thcor}
\ee
It is also useful to calculate the commutator
\be
\la [\hat{x}(t),\hat{x}(0)]\lb \equiv \mbox{Tr}[e^{-\beta H} [\hat{x}(t),\hat{x}(0)]]
=-\f{i}{\omega}\sin\omega t \ ,
\ee
which is actually independent of the temperature. This leads to the retarded Green function
\be
G_{R}=\la [\hat{x}(t),\hat{x}(0)]\lb_{R} \equiv \theta(t)\mbox{Tr}[e^{-\beta H}  [\hat{x}(t),\hat{x}(0)]]
=-\f{i}{\omega}\theta(t)\sin\omega t \ . \label{retcorh}
\ee

In the free scalar field theory (\ref{freeh}) with a mass $m$, if we work in the momentum
basis $k$, the calculations are identical to Harmonic oscillators
with the $k$-dependent frequency $\omega_k=\s{m^2+k^2}$. Therefore
the time ordered two point function is found to be (\ref{thpo}) as is clear from (\ref{thcor}).

\subsection{Mass Quench}

Now we would like to analyze the setup of the quantum quench due to the sudden change of frequency $\omega$ in
a harmonic oscillator (\ref{harmo}) as considered in \cite{CCd}.
Let us assume for $t<0$ the frequency is given by $\omega_0$ and the system
is in a thermal equilibrium at temperature $T_0=1/\beta_0$. At $t=0$, the frequency suddenly is changed into
$\omega$ by a certain external force and we are interested in physics when $t>0$.
Since the $\hat{x}$ and $\hat{p}$ are continuous at $t=0$, the creation and annihilation operators change
at $t=0$
from $(a_0,a_0^\dagger)$ to $(a,a^\dagger)$ via the following Bogoliubov transformation
\ba
&& a^\dagger=\cosh\zeta~ a_0^\dagger+\sinh\zeta~ a_0 \ ,\ \ \ \
a=\sinh\zeta~ a_0^\dagger+\cosh\zeta~ a_0 \ , \no
&& \cosh\zeta\equiv\f{1}{2}\left(\s{\f{\omega}{\omega_0}}+\s{\f{\omega_0}{\omega}}\right) \ .\label{anih}
\ea
Using (\ref{anih})
and (\ref{evol}), we can obtain the time-ordered two point function after the quench \cite{CCd}
\ba
&& T\la \hat{x}(t_1)\hat{x}(t_2)\lb\equiv T\left[\mbox{Tr}[e^{-\beta_0H_0}\hat{x}(t_1)\hat{x}(t_2)]\right] \no
&&\ \ \ \ =\f{1}{2\omega}e^{-i\omega|t_1-t_2|}+\left[\f{\omega_0}{4}\left(\f{1}{\omega^2}+\f{1}{\omega_0^2}
\right)\coth\f{\beta_0\omega_0}{2}-\f{1}{2\omega}\right]\cos\omega(t_1-t_2)\no
&&\ \ \ \ +\f{\omega_0}{4}\left(\f{1}{\omega^2_0}-\f{1}{\omega^2}\right)\coth\f{\beta_0\omega_0}{2}
\cos\omega(t_1+t_2) \ .\label{toh}
\ea
In section 5.1 we discusses the quantum quench in the free massive scalar field theory, where the mass is suddenly changed
from $m_0$ to $m$. In this case we can simply replace $\omega$ and $\omega_0$ with the $k$-dependent
frequency $\omega_k=\s{k^2+m^2}$
and $\omega_{0k}=\s{k^2+m_0^2}$, respectively.
This leads to the two point function (\ref{PropM}).

\end{document}